\documentclass[3p,12pt]{elsarticle}
\usepackage{graphics}
\usepackage{amsmath,amsthm,amsfonts,amssymb}
\usepackage{amssymb}

\journal{Physics Reports}
\begin{document}
\begin{frontmatter}

\title{The physics surrounding the Michelson-Morley experiment and a new \ae ther theory}

\author{Israel P\'erez}
\ead{cooguion@yahoo.com, iperez@mda.cinvestav.mx}
\address{Department of Applied Physics, CINVESTAV, M\'erida, Yucat\'an, M\'exico}

\begin{abstract}
From the customary view the Michelson-Morley experiment is used to expose the failure of the aether theory. The key point in this experiment is the \emph{fringe shift} of the interference pattern. Regularly, the fringe shift calculations are only presented from the perspective of the inertial frame where the one-way speed of light is anisotropic which gives a partial vision of the problem. In a spirit of revision of these facts we have meticulously analyzed the physics behind them. As a result, an angular effect which is based on Huyghens principle and plays a fundamental role in the reflection of light waves at moving mirrors is incorporated. Moreover, under the assumption of a null result in the experiment, on the one hand, the fringe shift conditions demand actual relativistic effects; on the other, it is confirmed that Maxwell's electrodynamics and Galilean relativity are incompatible formulations. From these two points at least three inertial theories follow: (1) the special theory of relativity (SR), (2) a new aether theory (NET) based on the Tangherlini transformations and (3) emission theories based on Ritz' modification of electrodynamics. A brief review of their physical content is presented and the problem of the aether detection as well as the propagation of light, within the context of SR and the NET, are discussed. Despite the overwhelming amount of evidences that apparently favors SR we claim that there are no strong reasons to refuse the aether which conceived as a continuous material medium, still stands up as a physical reality and could be physically associated with dark matter, the cosmic background radiation and the vacuum condensates of particle physics.
\end{abstract}


\end{frontmatter}

\section{Introductory survey}
\label{int}

During the second half of the XIX century, the theory of electrodynamics had been established by James Clerk Maxwell \cite{max1,max2,max3}. At that time, it was a fact that all known waves required a material medium to propagate and it was so natural to think that light waves could not be the exception. The medium was known as the \emph{luminiferous aether} which played the role of a privileged frame not only for the propagation of electromagnetic waves but for heat waves as well \cite{max1,whittaker}. As time went by the aether became the holy grail of physicists so that it was imperative to prove its existence. For this purpose, Maxwell suggested that experiments of order $v^2/c^2$ could be performed by measuring the velocity of the earth $v$ relative to the aether. In 1887 A. Michelson and E. Morely took the enterprise of carrying out an interferometric experiment \cite{michelson0,michelson1} to detect this motion by means of a fringe shift. Using \emph{ordinary kinematics} they thought that a considerable fringe shift would have revealed the existence of the light medium. However, the experimental outcome was almost null in disagreement with the calculations. Consequently, the experiment has been considered (among other similar experiments) as a clear proof of the non-existence of the aether and the isotropy of the speed of light for all inertial observers. Nevertheless, under a new vision, we shall explain that a null result must be assumed, on one hand, only for purposes of theorization and, on the other, to expose the inconsistency between ordinary electrodynamics and Galilean relativity (GR), from which does not follow that there is no light medium. We consider a new aether theory and make clear that only the two-way speed of light is constant for all inertial observers. Also, it shall be shown that when the experiment is realized in non-vacuum conditions the speed of the earth might be detected. We further present a new insight for a more intelligible theory based on the tenet that space is a material continuum.

\section{The rules of the game}
\label{poindep}

Before we start with our investigations, let us situate ourselves in the atmosphere of physics before 1905. First of all, we must recognize that most philosophers, mathematicians and physicists of the XIX century believed that space existed as the container of objects. They thought that if all objects were removed, space would continue to exist. This vision, as it is well known, epitomizes a Kantian or Newtonian notion of space. From this view, space and Euclidean geometry go side by side. Scientists also clearly distinguished space from matter. But one issue that they rarely discussed was the constitution of space; for most thinkers space itself existed as ``a priori intuition", as ``a relational thing", as ``a substance" \footnote{In these respects two philosophical currents, i.e. relationalism and substantivalism, still prevail.}. In this work, however, we shall adopt a distinct posture by borrowing an elemental principle from the philosophy of materialism which simply states that: \emph{Everything that exists must be made up of something which we shall call matter} \cite{perez5}. Therefore, we shall assume the position that if space is real and exists it must be made up of \emph{matter}. Based on this principle it is interesting to differentiate among the types of matter that we can find in the universe. Hence, in the second place, allow me to quote the notion of aether and field as posed by James Clerk Maxwell \cite{max2}: 
\begin{description}
  \item[(2)] \textsf{The mechanical difficulties, however, which are involved in the assumption of particles acting at a distance with forces which depend on their velocities are such as to prevent me from considering this theory as an ultimate one, though it may have been, and yet may be useful in leading to the coordination of phenomena.}
  
 \textsf{I have therefore preferred to seek an explanation of the fact in another direction, by supposing them to be produced by actions which go on the surrounding medium as well as in the excited bodies, and endeavouring to explain the action between distant bodies without assuming the existence of forces capable of acting directly at sensible distances.}
  \item[(3)] \textsf{The theory I propose may therefore be called a theory of the \emph{Electromagnetic Field}, because it has to do with the space in the neighborhood of the electric or magnetic bodies, and it may be called a \emph{Dynamical Theory}, because it assumes that \textbf{in that space there is matter in motion}, by which the observed electromagnetic phenomena are produced.}
  \item [(4)] \textsf{The electromagnetic field is that part of space which contains and surrounds bodies in electric or magnetic conditions.}
  
\textsf{It may be filled with any kind of matter, or we may endeaveour to render it empty of all \textbf{gross matter} as in the case of Geissler's tubes and other so-called vacua.}

\textsf{\textbf{There is always, however, enough of matter left to receive and transmit the undulations of light and heat}, and it is because the transmission of these radiations is not greatly altered when transparent bodies of measurable density are substituted for the so-called vacuum, that we are obliged to admit that the \textbf{undulations are those of an aethereal substance}, and not of the gross matter, the presence of which merely modifies in some way the motion of the aether.}

\textsf{We have therefore some reason to believe, from the phenomena of light and heat, that \textbf{there is an aethereal mediun filling space and permeating bodies}, capable of being set in motion and of transmitting that motion from one part to another, and communicating that motion to gross matter so as to heat it and affect it in various ways.}
\end{description}

From these paragraphs it is worth noticing the following points. First, for Maxwell, electromagnetic fields were dynamical states of ``aethereal matter". Second, for him space existed even if there were nothing (Newtonian notion). However, note that for Maxwell the conception of \emph{vacuum} was not a synonymous of \emph{empty space}, but only removal, from that region of space, of gross or ``macroscopic" matter (electrons, atoms, molecules, etc.); and such region, however, is still filled with aethereal matter. Therefore, if we follow the materialist line of thought we are allowed to surmise that space, vacuum and aether are the same thing. Third, we can distinguish at least two types of matter, macroscopic or gross matter, which in modern terminology corresponds to the set of particles of the standard model, and aethereal matter which we must still delve about its nature.

Thus, according to the beliefs of that time, we shall assume that electromagnetic waves travel in free aether (i.e., vacuum or space) and are governed by the wave equation:
\begin{equation}
\label{waveeq}
\triangle \Psi-\frac{1}{c^2}\frac{\partial^2 \Psi}{\partial t^2}=0,
\end{equation}
where $\Psi$ is a wave function (of an electric $\mathbf{E}$ or magnetic field $\mathbf{B}$), $\triangle=\nabla^2$ is the Laplacian operator, $c=1/\sqrt{\mu_0 \epsilon_0}$ is the wave velocity and $\mu_0$ and $\epsilon_0$ are the dielectric and magnetic constants, respectively. We conventionally propose that such equation was discovered by an observer at rest in the aether frame $S$. We now wonder what kind of law governs electromagnetic wave phenomena in another frame $S'$ which is in relative uniform motion with respect to $S$. 

In accordance with the above lines we must emphasize the following two statements: 
\begin{itemize}
  \item [a)] \emph{The aether is the material medium whereby electromagnetic disturbances propagate at isotropic speed $c$}.
  \item[b)] \emph{The source of the electromagnetic disturbances does not drag the aether with it}.
\end{itemize}
The latter statement implies that the speed of the electromagnetic waves remains constant when it is measured with respect to the aether and, therefore \emph{it is independent of the state of motion of the emitting body}. 

Finally, we must remark that as a consequence of the Galilean theorem of addition of velocities, physicists used to believe that there was no upper limit for the propagation speed of physical entities, so macroscopic objects (or the light source itself) could surpass the speed of light. This is not implicit in Maxwell's formulation of electrodynamics. And this is the reason why, in his article of 1904, Lorentz imposed the restriction that ``\emph{the speed of a system be less than that of the speed of light}" \cite{lorentz2}. Later it has been realized that the speed of light (assuming the aether at rest) represents a limiting speed for the propagation of all physical entities. Thus, to have an intuitive picture of light behavior it is healthy to assume that the speed of the light source cannot exceed the speed of the wave fronts.

\section{Fringe shift conditions}
\label{fringeshift}

One of the key points of the Michelson-Morley (MM) experiment is the fringe shift of the interference pattern produced by the superposition of the different waves. So, before we proceed further, first let us briefly recall the theory of interference \cite{born} which shall play one of the fundamental roles in this work.

\begin{figure}[htp]
\begin{center}
\includegraphics[width=4cm]{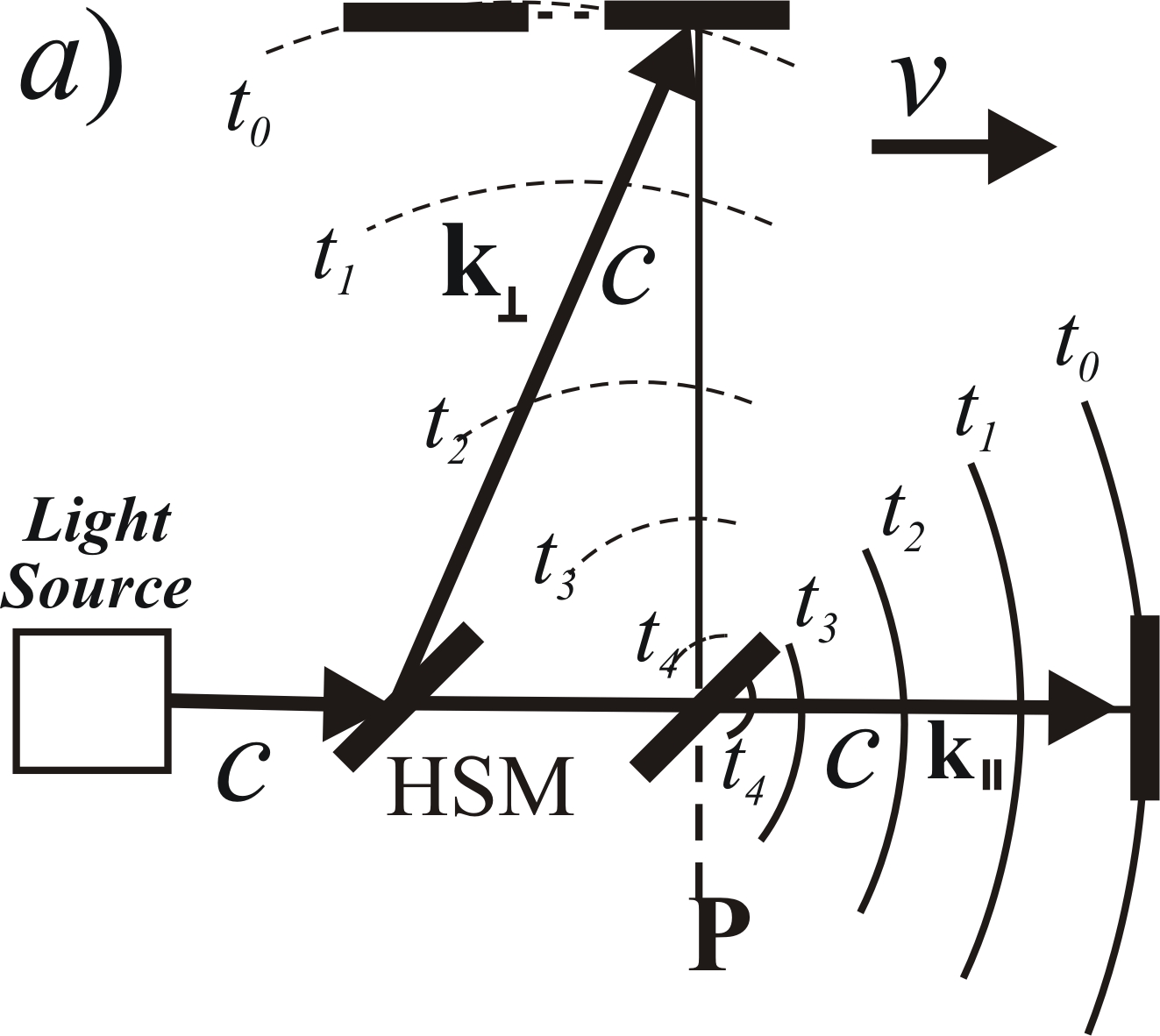} \hspace{2cm} \includegraphics[width=4.5cm]{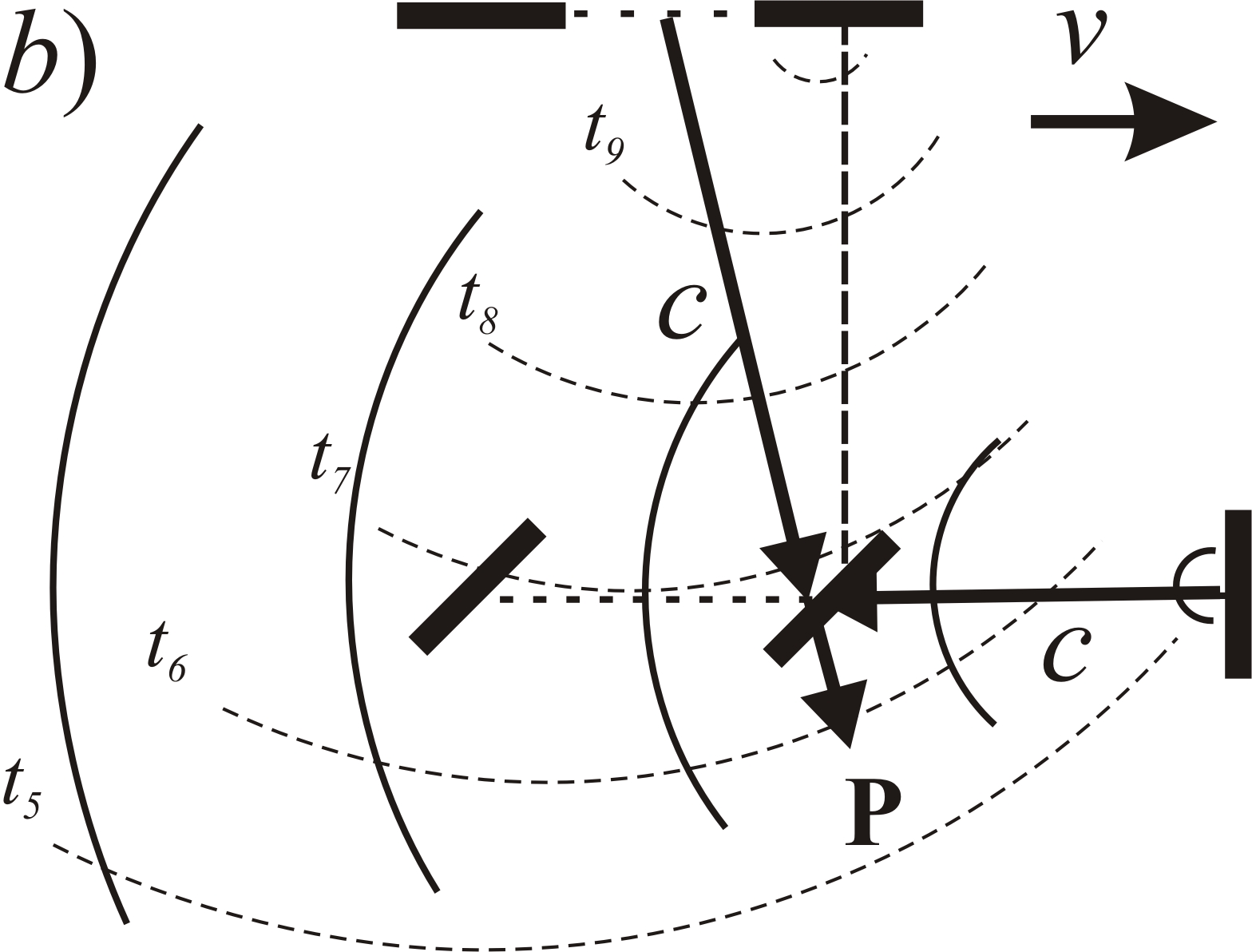}
\caption{The Michelson-Morley experiment as seen from the aether frame. ($a$) Forward advance of light waves. ($b$) Backward advance. Arrows represent the propagation vectors of the four wave fronts. Solid arcs are for longitudinal wave motion and dashed arcs for transversal wave motion.}
\label{fisk}
\end{center}
\end{figure}

Let two solutions of the wave equation \eqref{waveeq} be polarized-monochromatic plane waves in the transversal and longitudinal directions of the form (see Fig. \ref{fisk}):
\begin{eqnarray}
\label{elec}
\Psi_{\|} & = & \frac{1}{2}[\mathbf{A}e^{-i\omega t}+\mathbf{A}^*e^{i\omega t}], \nonumber\\
\Psi_{\bot} & = & \frac{1}{2}[\mathbf{B}e^{-i\omega t}+\mathbf{B}^*e^{i\omega t}].
\end{eqnarray}
$\mathbf{A}$ and $\mathbf{B}$ are complex vectors with the components
\begin{eqnarray}
\label{elecampl}
A_x = ae^{i\mathbf{k_{\|}}\cdot \mathbf{r}-\varphi_1}; \qquad A_y = ae^{i\mathbf{k_{\|}}\cdot \mathbf{r}-\varphi_2}; \nonumber \\
B_x = ae^{i\mathbf{k_{\bot}}\cdot \mathbf{r}-\xi_1}; \qquad B_y = ae^{i\mathbf{k_{\bot}}\cdot \mathbf{r}-\xi_2};
\end{eqnarray}
where $\mathbf{k_{\|}}$ and $\mathbf{k_{\bot}}$ are the propagation vectors in the longitudinal and transversal directions, respectively, such that $k\equiv 2\pi/\lambda=|\mathbf{k_{\|}}|=|\mathbf{k_{\bot}}|$ is the wave number and $\lambda$ is the wavelength, $\mathbf{r}$ is the position vector of the wave front and, $a$ is a constant real amplitude; $\varphi_j$ and $\xi_j$ (with $j=1,2$) are arbitrary phase angles, respectively. When the two waves superposed at a given point P, the total field is given by $\mathcal{R}\{\Psi\}=\mathcal{R}\{\Psi_{\|}+\Psi_{\bot}\}$, where $\mathcal{R}$ denotes the real part of the field. The intensity is proportional to the time average of the field squared, namely:
\begin{equation}
\label{maxin}
I\propto \langle \Psi^2\rangle=\langle \Psi_{\|}^2\rangle+\langle \Psi_{\bot}^2\rangle+\langle 2 \Psi_{\|} \cdot  \Psi_{\bot}\rangle.
\end{equation}
The first two terms are the intensities of the individual fields whereas the last term represents the interference. If the experimental conditions are such that the phase angles of the corresponding components are $\xi_j=\varphi_j$ then the interference term is found to be
\begin{equation}
\label{averint}
I_{12}\equiv \langle 2 \Psi_{\|} \cdot  \Psi_{\bot}\rangle=2a^2\cos(\delta),
\end{equation}
where we have introduced the phase difference
\begin{equation}
\label{eqpha}
\delta\equiv \mathbf{r}\cdot (\mathbf{k_{\|}}-\mathbf{k_{\bot}}).
\end{equation}
From equations \eqref{elec}-\eqref{averint} it follows that
\begin{eqnarray}
\label{totint}
I_1=I_2=\langle \Psi_{\|}^2\rangle&=&\langle \Psi_{\bot}^2\rangle=a^2; \qquad \nonumber \\
I  =  I_1+I_2+I_{12} &=& 4I_1\cos^2(\delta/2).
\end{eqnarray}

Expression \eqref{totint} also applies to spherical waves of the form
\begin{equation}
\label{elec2}
\Psi_{\|}  =  \frac{a}{r_\|}e^{i(kr_{\|}-\omega t)}, \quad \Psi_{\bot} =  \frac{a}{r_\bot}e^{i(kr_{\bot}-\omega t)},
\end{equation}
where $r_{\|}$ and $r_{\bot}$ are the radii of the spherical wave fronts that superposed at P, i.e., they specify the physical path lengths (PPL) $s_{\|}$ and $s_{\bot}$ of the waves from the source(s) to P. In such case the phase difference is given by
\begin{equation}
\label{eqpha2}
\delta=k(r_{\|}-r_{\bot})=k\delta s
\end{equation}
where $\delta s=s_{\|}-s_{\bot}$ is the difference between the PPL. From the preceding formulation it is clear that \emph{an interference pattern is only produced when there exists a phase difference between the waves}. If this requirement is satisfied, it follows from Eq. \eqref{totint} that a maxima of intensity is obtained when $\delta /2=\pi N$ ($N=0,1,\dots$), hence by Eq. \eqref{eqpha2} (or Eq. \eqref{eqpha}) the number of fringes is given by
\begin{equation}
\label{fringe1}
N=\frac{\delta s}{\lambda}=\nu \delta t.
\end{equation}

The theory of interference we have reviewed above is in accordance with the experimental conditions of the MM experiment. Traditionally, plane waves or rays are used to calculate the fringe shift, however, here we encounter a conceptual difficulty. For a stationary situation this procedure appears satisfactory, but when the source is in motion it leaves much to be desired. My view may be expressed by saying that rays are just idealizations to help us with the direction of energy flow and that an actual light beam has a finite cross section. 
Moreover, plane waves are just approximations for the description of spherical waves emitted from infinite distant sources. And finally, the Doppler effect is harder to visualize if we follow this path. Consequently, a more realistic situation is accomplished, if we bear in mind that spherical waves are emitted from the sources. This position is founded, firstly, upon the fact that spherical waves diverge in all directions from the origin as function of their radius similarly as light beams increase their cross section. Secondly, spherical waves satisfy the inverse square law for the intensity, as can be verified from Eqs. \eqref{totint} and \eqref{elec2}, whereas plane waves do not \cite{born}. And, thirdly, the Doppler shift is easier to visualize with spherical waves. Based on these statements, a mental scheme of light behavior for the MM experiment is presented in Fig. \ref{fisk}. For the sake of illustration, the extension of the waves fronts have been accentuated. The interferometer is shown in motion ($v<c$) relative to the aether where Doppler-shifted waves with their respective propagation vectors are depicted as well.

On the other hand, note that in equation \eqref{fringe1} we have made used of the ansatz $c=\lambda \nu$ to interchange to a time domain with $\delta t=t_{\|}-t_{\bot}$, where $t_{\|}$ and $t_{\bot}$ are the times elapsed by the waves in the longitudinal and transversal journeys, respectively. This change was possible because in the solutions \eqref{elec} or \eqref{elec2}, $\omega$ and $k$ have a linear relation, i.e., the group and the phase velocity $V\equiv \partial \omega/\partial k=\omega/k=c$ are the same in all directions. However, before the discovery of the Lorentz transformations the only known transformations relating two inertial frames, moving with relative velocity $\mathbf{v}$, were the Galilean transformations:
\begin{equation}
\label{galtrans}
\mathbf{r'}=\mathbf{r}-\mathbf{v}t; \qquad t'=t.
\end{equation}
As a consequence physicists used to believe, in analogy with the speed of sound, that the velocity of light could acquire different numerical values in inertial frames in motion relative to the aether and, thus, the wave numbers $k$'s or the frequencies $\omega$'s for each of the waves would not take on, in general, the same values. This can be easily shown by applying these transformations to Eq. \eqref{waveeq}, from which we will obtain an equation in $S'$ of the form \cite{jackson}:
\begin{equation}
\label{waveeqin}
\biggl[\mathbf{\triangle}-\frac{1}{c^2}\frac{\partial^2}{\partial t^2}-\frac{2}{c^2}\mathbf{v} \cdot \mathbf{\nabla} \frac{\partial}{\partial t}- \frac{1}{c^2} \mathbf{v} \cdot \mathbf{\nabla} \mathbf{v} \cdot \mathbf{\nabla} \biggr]\Psi=0.
\end{equation}
A one-dimensional solution of the this equation may be given by
\begin{equation}
\label{wavesol}
\Psi (x,t)=ae^{i(kx-\omega t)},
\end{equation}
where $x$ is the position of the wave front and $a$ is the amplitude. Considering the motion of the frames in the $x$ direction, the velocities $V_{\pm}$ are found out by replacing the solution \eqref{wavesol} into the wave equation \eqref{waveeqin}, from which we will obtain
\begin{equation}
\label{speeds}
V_{\pm}=\frac{\omega}{k_{\pm}}=c\mp v,
\end{equation}
where the subindices $\pm$ are assigned for the forward and backward advances, respectively; and $v=|\mathbf{v}|$. 

Therefore, if we allow the velocities $V_i$ (with $i=1,2,3,4$) of the four waves to vary, the phase difference \eqref{eqpha2} (or \eqref{eqpha}) might depend, in general, not only on the optical paths lengths OPL= $\int n(s_i) ds_{i}$ transversed by each of the waves but also on their velocities. Because of all waves travel in free aether, the refractive index $n=constant$ is the same for all waves, then the OPL reduces to the calculation of the individual PPL $s_i=\int ds_i$, and since $V_i=\omega/k_i$ a more general expression for the phase difference is given by
\begin{equation}
\label{fringe2}
\delta= k_1s_1+k_2s_2-(k_{3}s_{3}+k_{4}s_4)=\omega \delta t,
\end{equation} 
where
\begin{equation}
\label{nfringe}
\delta t=\frac{V_3V_4(s_2V_1+V_2s_1)-V_1V_2(V_3s_4+V_4s_3)}{V_1V_2V_3V_4}. 
\end{equation}
Here we have considered the indices 1 and 2 for the longitudinal direction, forward and backward advance, respectively; and, 3 and 4 for the transversal direction forward and backward advance, respectively.

Let us see two important cases. (I) In the case that the velocities are all fixed, e.g., $V_i=c$, the foregoing equation becomes the traditional one (cf. with Eq. \eqref{eqpha2}), namely
 \begin{equation}
\label{nfringe2}
\delta t=\frac{\delta s}{c},
\end{equation}
with $s_{\|}=s_1+s_2$ and $s_{\bot}=s_3+s_4$. (II) In the case that the PPL are all fixed, e.g., $s_i=\zeta$ with $\zeta$ a constant, Eq. \eqref{nfringe} becomes
 \begin{equation}
\label{nfringe3}
\delta t=\zeta\frac{\delta V}{V_1V_2V_3V_4},
\end{equation}
with $\delta V\equiv V_3V_4(V_1+V_2)-V_1V_2(V_3+V_4)$. We shall name this as the ``difference between the velocities". In plain words, assuming $\omega=constant$, from these cases we can deduce that a fringe shift $\delta$ might be induced when the next conditions are satisfied:
\begin{enumerate}
  \item \textsf{If $\delta V=$constant, then $\delta t\propto \delta s$.} 
  \item \textsf{If $\delta s=$ constant, then $\delta t\propto \delta V$.} 
\end{enumerate} 
In addition, if both $\delta s$ and $\delta V$ are not simultaneously constant, then $\delta t$ is given by Eq. \eqref{nfringe}. Finally, if $\omega$ is not constant we use Eq. \eqref{fringe2}. We can summarize all of this by stating that $\delta\equiv \delta(\delta s,\delta V,\omega)$, thus a fringe shift is induced only if $d\delta/dt \ne 0$.

\section{The effective angle}
\label{misang}

In the original experiments \cite{michelson0,michelson1}, Michelson and Morley set the angle of the half-silvered mirror (HSM) to $\varphi_0=\pi/4$ rad so that the transversal rays could form a right angle relative to the abscissa. In Fig. \ref{fisk}, however, the transversal rays, even those that superposed at P, are not perpendicular to the abscissa. What is the physical cause of this angular difference? Based on Huyghens' principle (HP) the answer was given by the authors themselves in the supplement of their article of 1887. This rests upon the following fact which is not exclusive of this experiment.
\begin{figure}[htp]
\begin{center}
\includegraphics[width=5cm]{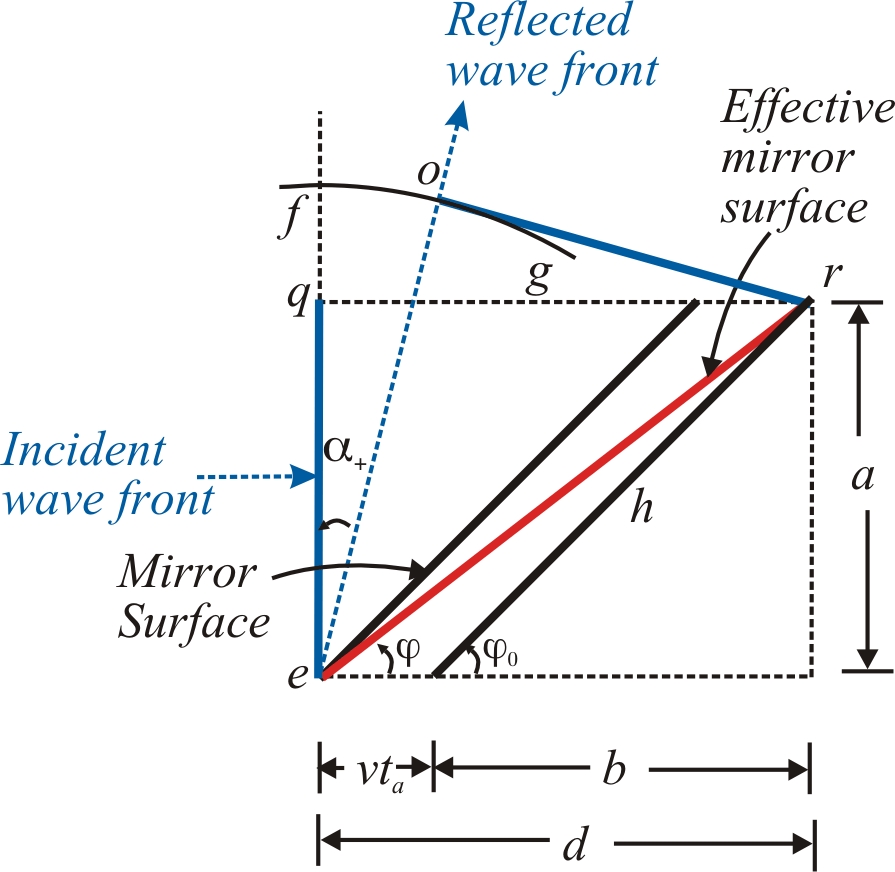}
\caption{(Color online) Illustration of the effective angle for the transversal wave motion as the HSM moves to the right. The reflected wave front leaves the HSM with an angle $\alpha_+$.}
\label{mirror}
\end{center}
\end{figure}

In Fig. \ref{mirror} a zoom of the HSM is shown as it moves to the right. If we take into consideration that a wave front has a finite extension $a$, then at a given time the lower edge of the wave front interacts with the HSM but the upper one still needs some more time $t_a$ to reach the corresponding part of the HSM. During this time the HSM has moved the distance $vt_a$, thus, in reality, the wave front does not ``see" the angle $\varphi_0$ but an effective angle $\varphi$ which is velocity dependent. This effect is also present for the backward-longitudinal (BL) direction of the wave fronts when these interact with the HSM; so that both wave fronts, transversal and longitudinal, superposed at the point P.

With the aid of the Fig. the effective angle is simply $\tan{\varphi}=a/d$, where $a=h\sin \varphi_0$ and $d$ is the distance that the upper edge of the wave front has to travel to reach the HSM during the time $t_a$. This time is obtained as follows:
\begin{equation}
\label{d1}
d=ct_a=vt_a+b; \qquad t_a=\frac{b}{c-v}=\frac{h \cos \varphi_0}{c-v};
\end{equation}
hence $d=a\cot \varphi_0/(1-\beta)$, where $\beta=v/c<1$. With this at hand we have
\begin{equation}
\label{d2}
\tan \varphi=(1-\beta)\tan \varphi_0.
\end{equation}
To determine the new wave front we draw the arc $fg$ of radius $d$ with $e$ as a center. The wave front is the tangent from $r$ to the arc $fg$; and the new direction will be the normal to this tangent from $e$. This direction is given by the angles $\alpha_{\pm}$ of the wave-normal, where the signs $\pm$ stand for the forward-transversal (FT) direction and BL direction, respectively. For the sake of illustration we shall calculate $\alpha_+$, the calculation of $\alpha_-$ is straightforward. Because of the triangles $eqr$ and $eor$ are equal, it follows that $\varphi=\pi/4-\alpha_+/2$ and 
\begin{equation}
\label{compl}
\tan \varphi= \tan\biggl( \frac{\pi}{4}-\frac{\alpha_+}{2}\biggr)=\frac{1-\tan \frac{\alpha_+}{2}}{1+\tan \frac{\alpha_+}{2}}.
\end{equation}
A similar expression is found for the BL wave front; then solving for $\alpha_{\pm}$ from Eqs. \eqref{d2} and \eqref{compl} after a little algebra we found that
\begin{equation}
\label{efangle}
\tan \frac{\alpha_{\pm}}{2}=\pm\frac{1-(1\mp \beta)\tan \varphi_0}{1+(1\mp \beta)\tan \varphi_0}.
\end{equation}
In our case $\varphi_0=\pi/4$, hence 
\begin{equation}
\label{alo}
\tan{\frac{\alpha_\pm}{2}}=\frac{\beta}{2\mp \beta};
\end{equation}
$\alpha_+$ is also the angle at which the FT wave-normal reaches the upper mirror of the transversal arm and, subsequently, it is reflected with the same angle towards the HSM. Whereas $\alpha_-$ is the angle at which the BL wave-normal arrives at the point P after interaction with the HSM. This angular phenomenon will play one of the fundamental roles in this work. 

\section{Fringe shift calculations}
\label{clasrea}
What we have seen in the previous sections was the preamble to fully understand the basic notions of light behavior based only on optics theory. Once that we have elucidated these nuances we can proceed to analyze the different calculations to determine the fringe shift.

\subsection{Approach A}
\label{inta}
First of all, remember that the arms of a Michelson interferometer are fixed and of equal length, say, $l_0$. Thus, if the apparatus is at rest in $S$, we have
\begin{eqnarray}
s_{\|0}=s_{\bot 0}=2l_0, \nonumber  \\
t_{\|0}=t_{\bot0}=\frac{2l_0}{c},
\end{eqnarray}
hence, $\delta V=0$ and $\delta s=c\delta t=0$, and there is no fringe shift. But if the apparatus were set in motion the problems in the interpretation arise\footnote{For more general treatments of this perspective see the works of Articolo \cite{articolo} and Mazur \cite{mazur}.}. Let us first analyze the experiment as judged by an observer in $S$ (see Fig. \ref{fisk}).  

\begin{figure}[htp]
\begin{center}
\includegraphics[width=7cm]{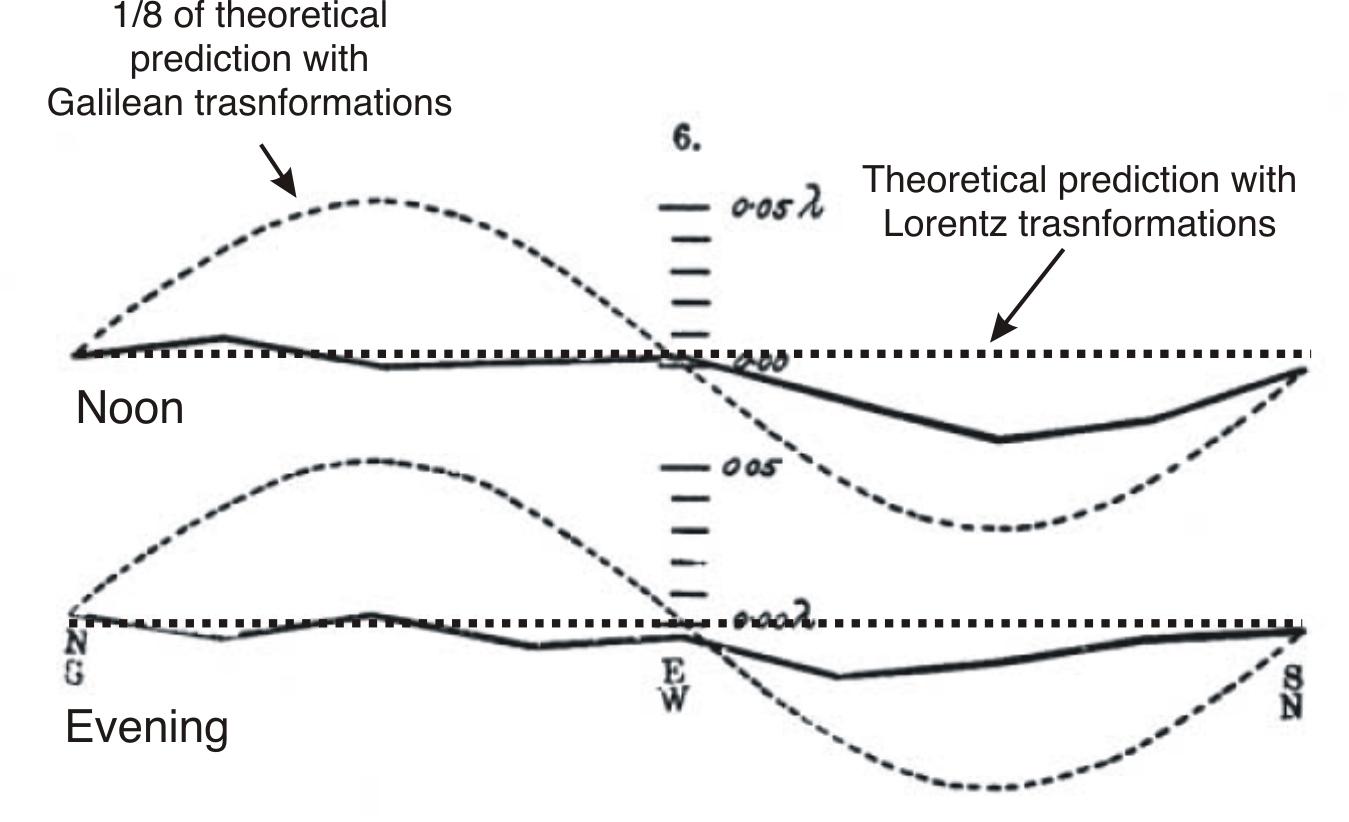}
\caption{The observed fringe shift. Adapted from the results of the Michelson-Morley Experiment of 1887: The upper is the curve for observations at noon, and the lower for that of the evening observations. Experimental data in solid lines. Dashed lines represent one eighth of the expected displacements using ordinary kinematics. Dotted lines shows the theoretical prediction of SR or the new ather theory in vacuum conditions.}
\label{michelson}
\end{center}
\end{figure}

Suppose that the earth frame $S'$ carrying the interferometer moves with respect to the $S$ frame at the velocity $\mathbf{v}=v\,\mathbf{\hat{x}}$ ($v<c$) and that the origin of coordinates of both frames coincide at $t=t'=0$. Since we are interested in the physical interpretation we shall assume, for simplicity, that the motion of  the arm in the longitudinal direction is parallel to $\mathbf{v}$. Then, bearing in mind our above statements, we see that $\delta V=0$ and, therefore, condition 1 given above applies. Let us then calculate the shift $\delta s$. For the round trip in the transversal direction, the wave fronts travel the total distance
\begin{equation}
\label{haz1}
s_{\bot}\equiv ct_{\bot}=\sqrt{(vt_{\bot})^2+(2l_0)^2}.
\end{equation}
On solving for the time yields
\begin{equation}
\label{ttrans}
t_{\bot}=\frac{2l_0}{c}\gamma,  \quad \textrm{and}\quad s_{\bot}=ct_{\bot},
\end{equation}
where $\gamma=1/\sqrt{1-\beta^2}$. For the longitudinal direction the wave fronts travel the distances
\begin{equation}
\label{s1}
s_{1} \equiv ct_1= l_0+vt_1, \qquad s_{2}\equiv ct_2= l_0-vt_2,
\end{equation}
for the forward and backward advance, respectively; with $t_i,\; (i=1,2)$ the times elapsed for each journey. Solving for the times from the previous equations we find that
\begin{equation}
\label{t1}
t_1=\frac{l_0}{c-v}; \qquad t_2=\frac{l_0}{c+v}.
\end{equation}
So, considering that $t_\|=t_1+t_2$, we obtain
\begin{equation}
\label{tparal}
t_\|=\frac{2l_0}{c} \gamma^2, \quad \textrm{and}\quad s_{\|}=ct_{\|}.
\end{equation}
Subtracting Eq. \eqref{ttrans} from Eq. \eqref{tparal} and expanding to a first approximation, there is a maximum time shift given by
\begin{equation}
\label{deltat}
\delta t\approx \frac{2l_0}{c}\beta^2.
\end{equation}
Here the factor $2$ appears as consequence of a $\pi/2$ rotation. It is also evident that $s_{\|}>s_{\bot}$, as a result, \emph{the fringe shift is only function of the variations of the difference of the PPL} $\delta s(v)=c\delta t$. At this point we must remark that as long as we regard the earth as an inertial frame no fringe shift is expected. This fact supports the principle of relativity as enunciated by Poincar\'e who stated as a general law of nature the ``impossibility to detect the absolute motion of the Earth by experiment" \cite{poincare2,scribner}. The equation is telling us that the experiment is sensitive to the variations of speed, and since the earth is a non-inertial frame we shall expect variations of the fringes (see Fig. \ref{michelson}). This is what Michelson and Morley were seeking for.

Next, the observer in $S$ must calculate the laws of physics that occur on earth. But since the earth is not an inertial frame, being strict, the Galilean transformations do not apply in this case. Instead, the correct path to tackle this problem is using transformations that relate inertial frames with non-inertial frames (see for example \cite{wu,ashworth,hsu2,hsu1}). However, to a first approximation the inertial approach could give us an idea of the physics. Hence, the application of the Galilean transformations \eqref{galtrans} will lead us to find that the velocities of the energy flow for the four waves in the frame $S'$ take on the values:
\begin{equation}
\label{trans}
\mathbf{c}'_{\| \pm} =\pm(c\mp v)\; \mathbf{\hat{x}}'; \quad \mathbf{c'_{\bot \pm}}=\pm \sqrt{c^2-v^2} \;\mathbf{\hat{y}}'.
\end{equation}
The $\pm$ signs are assigned for the forward and backward advances, respectively. Whilst the times and frequencies remain the same, that is, $t'=t$ and $\nu'=\nu$. 

\subsection{Approach B}
\label{intb}
\begin{figure}[htp]
\begin{center}
\includegraphics[width=5.5cm]{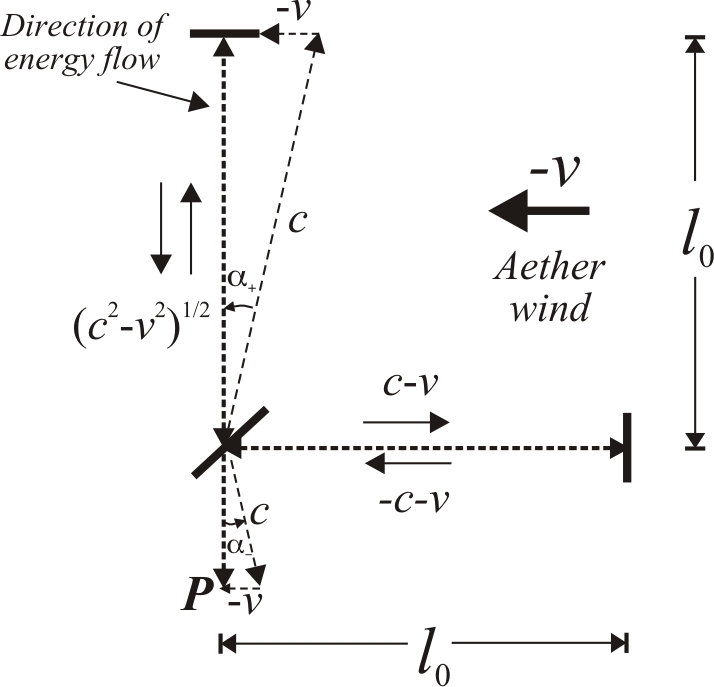}
\caption{Michelson-Morley configuration as seen from the earth frame. Using Galilean relativity, the one-way speed of light waves as calculated by an observer in the frame S are anisotropic. Due to the angular effect that occurs in the mirrors the energy flow of the transversal waves is perpendicular to the motion.}
\label{mm1}
\end{center}
\end{figure}

But now, as judged from $S'$, the aether is passing by with velocity $\mathbf{v}'=-v \, \mathbf{\hat{x}}'$ (see Fig. \ref{mm1}) and if Galilean transformations were correct, the speed of light, as predicted by the observer in $S$ from approach A, would take on the velocities given in equation \eqref{trans}.
Then, for  the parallel direction the wave front travels the distances
\begin{equation}
\label{espac}
s'_1=l_0=t'_{1}(c- v),  \qquad s'_2=l_0=t'_{2}(c+v),
\end{equation}
for the forward and backward advances, respectively. On solving for the times we obtain
\begin{equation}
\label{t2}
 t'_{1}=\frac{l_0}{c- v}; \qquad  t'_{2}=\frac{l_0}{c+ v}.
\end{equation}
By adding the times we would evidently arrive at a time expression numerically identical to that of Eq. \eqref{tparal}. However, notice that in such case $s'_{\|}=2l_0\neq ct'_{\|}=s_{\|}=ct_{\|}$.

To complete the calculation, perhaps, we may take the transversal distance from expression \eqref{haz1} mixing, however, the interpretation of both frames. But strictly speaking, the discussion of section \ref{misang} revealed that light is actually reflected with an angle $\alpha_+$. Then, as seen from $S$, the direction of energy flow follows a diagonal towards the transversal mirror, but this same process, as judged from $S'$, is perceived completely vertical due to the aether drift with unit vector $\mathbf{\hat{m}'}=(c\mathbf{\hat{k}_{\bot}}-\mathbf{v})/|c\mathbf{\hat{k}_{\bot}}-\mathbf{v}|$. In such situation, the effective speed of energy flow is given simply by $c'_{\bot}=\sqrt{c^2-v^2}$, hence the transversal distances are
\begin{equation}
\label{abc}
s'_{3}=s'_{4}=l_0=c'_{\bot}\frac{t'_{\bot}}{2},
\end{equation}
again in agreement with equation \eqref{ttrans}. Also note that $s_{\bot} \neq s'_{\bot}=2l_0$. The combination of the preceding formulas yields
\begin{equation}
\label{deltatprim}
\delta t'\approx \frac{2l_0}{c}\beta^2.
\end{equation}
This equation could have been also obtained by the use of Eq. \eqref{nfringe3}, with $\zeta=l_0$, just replacing the corresponding velocities. Note that this result is numerically identical to equation \eqref{deltat}, however, in the former case, the speed of light is isotropic, therefore \emph{the physical interpretation of these equations is not the same; but it depends on the observer's frame}. Since $\delta s'=s'_{\|}-s'_{\bot}=0$ and $\delta V'(v) \neq const.$, condition 2 given above states that the physical interpretation is that \emph{the fringe shift is caused by the variations of the difference of the wave velocities as the interferometer moves through the aether}. This is the conventional interpretation that most physicists have accepted during the last hundred years, but we must not deceive ourselves by this line of thought, for such interpretation is fictitious since there is only one absolute speed of light. Hence, if we wish to be unbiassed we should not disregard the most relevant interpretation, that is, that of approach A.

\subsection{Approach C}
\label{intc}
But even more, following the \emph{theorem of addition of velocities} and the notion of rays some physicists reasoned that, as judged again from $S$, if the earth is moving with speed $v$ and the speed of light with respect to the source is $c$, then the absolute velocities of the waves for each direction would acquire the following values:
\begin{equation}
\label{threevel}
\mathbf{c}_{\| \pm}=\pm(c\pm v)\; \mathbf{\hat{x}}, \qquad \mathbf{c}_{\bot \pm}=v \;\mathbf{\hat{x}}\pm c\; \mathbf{\hat{y}}.
\end{equation}
However, this formulation contradicts our above statements because, in $S$, the velocity of light would depend on the velocity $v$ of the light source. Instead, this fact might imply that the \emph{aether is dragged along with the light source}. From this perspective it is obvious that since the very beginning we are facing the dilemma: is the wave equation \eqref{waveeq} describing correctly wave phenomena? or, is GR flawed? 

Besides, the dragging-aether hypothesis will lead us to the next formulae when judged from a stationary observer who is not dragged \footnote{In emission theories this frame is called the\emph{ Master Frame \cite{cyrenica}.}}. Let us see what we find. 

The distances $s_{\pm}$ traveled by the wave front in the direction of motion will be
\begin{equation}
\label{smas}
s_{\pm} \equiv (c\pm v )\tau_{\pm}= l_0\pm v\tau_{\pm},
\end{equation}
with $\tau_{\pm}$ the times for the forward and backward advance, respectively. Note that since, in this case, the velocities are faster (or slower) than in approach A, the times $\tau_{\pm}$ do not correspond with their counterparts $t_{i}$, that is, $t_{i}>\tau_{\pm}$, respectively. Upon solving for the total time $\tau_{\|}=\tau_{+}+\tau_{-}$, we obtain from the previous equations that
\begin{equation}
\label{ttimes}
\tau_{\|}=\frac{2l_0}{c}.
\end{equation}
And for the transversal wave we have the distance
\begin{equation}
\label{transmas}
s_{\bot}\equiv c_{\bot}\tau_{\bot}= \sqrt{(v\tau_{\bot})^2+(2l_0)^2},
\end{equation}
which on solving for $\tau_{\bot}$ we also arrive at $\tau_{\bot}=\tau_{\|}$ and therefore $\delta \tau=0$. It has to be noted that in this case $\delta V(v)\ne const.$ and $\delta s(v)\ne const.$, however, using Eq. \eqref{nfringe} we can easily reaffirm that $\delta \tau=0$, so there is no change in the interference pattern.

Now if the aether were dragged by the source there would be no aether wind in $S'$ and the speed of the waves would be $c$ in all directions. This can be easily shown by applying the Galilean transformations to Eq. \eqref{threevel}. Accordingly, it follows that $2l_0=c\tau'_{\|}=c\tau'_{\bot}$; hence $\tau'_{\|}=\tau'_{\bot}$, $\delta V'=0$ and $\delta s'=0$, and again, there is no fringe shift.

The preceding interpretation is the same to that put forward by George Stokes \cite{stokes}. And it is also kinematically equivalent to the so-called emission theories (ET) of Ritz \cite{pauli,ritz,fox2} and Tolman \cite{tolman0,tolman}, in which the behaviour of electromagnetic radiation is ballistic.

\subsection{Experimental conclusions based on fringe shift conditions}

As shown in Fig. \ref{michelson} the fringe shift of the original experiment was not really zero \cite{michelson1}. This may be so, in first place, because the experiment was conducted at normal conditions, i.e., not in vacuum. And in second place, as we discussed above, because acceleration effects contribute also to the fringe shift. In spite of this, modern experiments  \cite{kennedy,miller,miller2,shankland,jaseja,brillet,riis,hils,muller,muller1,chen,herrmann,antonini,wolf} have shown that the experimental data apparently tends to such value. Thus, for the purposes of theorization we shall assume that it is zero. Then, one would ask: what conclusions can be drawn?   

In approach A we have calculated that when the interferometer is in motion $\delta s (v)=2l_0\beta^2\ne const.$ and $\delta V=0=const.$, however, our null assumption contradicts these calculations, it rather suggests that one of the next three cases must be true: 
\begin{itemize}
  \item [(\textbf{*})] $\delta s\ne const.$ and $\delta V\ne const.$
  \item [(\textbf{\dag})] $\delta s=const.$ and $\delta V= const.$
  \item [(\textbf{\ddag})]  $\delta s= const.$ and $\delta V\ne const.$
  \end{itemize}
Each of these results will lead us to distinct theoretical systems and interpretations of the experiment. However, we will briefly focus on the first two items, the third one is outside the scope of the present investigation.
 
\section{Aether dragging and emission theories}

If item (\textbf{*}) were correct, we are in the realm of approach C as seen from $S$. In this case the following physical interpretation arises:
\begin{itemize}
  \item [(C.i)] The aether is dragged by the light source.
\end{itemize}
In contrast to SR where Maxwell's electrodynamics is maintained intact and GR is modified, in this case GR is correct but the whole of electrodynamics must be reformulated. Item C.i is nothing but the version of George Stokes \cite{stokes}. However, on one hand, at the age of Stokes there was not a complete formulation of electrodynamics. Maxwell partially resolved this issue assuming a coordinate system at rest with respect to the aether where his equations were valid and, thus a constant speed of light results as long as the source is at rest. But he did not explore the case when the light source is moved or when the aether is dragged (see for instance \cite{whittaker,ohara,hunt}. Later Lorentz assumed the aether at rest and developed the theory of electrons \cite{lorentz3} on this basis. 

This interpretation also alludes to the emission theories (ET) of Ritz et al. \cite{cyrenica,pauli,ritz,tolman0,tolman}. In these theories the speed of light is always $c$ with respect to the light source only. So Ritz kept the two homogenous Maxwell's equations intact but modified the source equations so that the scalar and vector potentials acquire the next form \cite{pauli}:
\begin{eqnarray}
\label{elecmod}
\Phi(P,t)& = & \int \frac{\rho dV_{P'}}{[r_{PP'}]_{t'=t-[r/(c+v_r)]}}; \nonumber \\  
\mathbf{A}(P,t) &= &\int \frac{(1/c)\rho \mathbf{v} dV_{P'}}{[r_{PP'}]_{t'=t-[r/(c+v_r)]}}.
\end{eqnarray}
Such modification avoids the inconsistency of the ordinary theory that results in the advance solution for radiation. The adherents of the ET argued that the invariance or covariance of physical laws are not  crucial requirements for physics. In fact, for quantities of first order there is no difference between ET and relativistic optics, provided one deals with closed paths. The latter condition is almost always (possibly always) satisfied, since by their very nature, all electromagnetic phenomena are circuital. Moreover, the electron in ET must adopt a different meaning. Actually, in order for ET to be consistent, Pauli pointed out that the whole of electrodynamics must be re-conceptualized anew \cite{pauli}. 

By assuming a ballistic behavior of light, ET can explain interferometric experiments like the MM experiment, the Fizeau experiment \cite{fizeau1,fizeau3,michelson2} and the Kennedy-Thorndike experiment \cite{cyrenica}. They also explain the stellar aberration of light, reflection, refraction, the Doppler effect, the effect of acceleration, Hubble's law for distant nebulae, etc. However, experiments about the constancy of the speed of light from moving sources \cite{jaseja,fox1,alvager} and the mass-energy relation \cite{bertozzi} constitute strong evidences against ET \footnote{See also the work of Fox \cite{fox2}.}.

In addition, based on the Galilean theorem of addition of velocities, at first glance, ET seem to contradict causality because there is no limit to the propagation of entities. However, if electromagnetic or gravitational phenomena are circuital they may constrain the speed of physical entities.

To end this section, it is worth mentioning that during the past hundred years most physicists have devoted their efforts to the investigation of relativistic theories, as a result, ET are relatively still in an initial stage of development. For this reason they cannot yet be excluded from physics.
 
\section{FitzGerald-Larmor-Lorentz-Poincar\'e vs Einstein-Minkowski}  
\label{reltheo}

If we still wish to keep up our statements of section \ref{poindep} and Maxwell's electrodynamics, then item (\textbf{\dag}) is our only choice. On looking for a rational explanation, the term $\delta s=const.=0$ demands that the difference of the PPL remains unaffected just as if the interferometer were at rest, \emph{therefore we must conclude that the dimensions of the arms must have altered} by a given factor. 

\subsection{Lorentz-FitzGerald contraction}
In our calculations for the fringe shift we have introduced the length $l_0$ taking for granted that \emph{physical objects behave as rigid bodies} \cite{pauli}. However, we all know that this is just an idealization to simplify the calculations and that in reality all objects undergo deformation when they are subjected to forces or temperature changes. Based on these facts, Lorentz knew that objects were made up of material particles (atoms and electrons) and glued by electric forces. And because of the interferometer is an object already in motion with respect to the aether, he thought that the aether must have been exerting forces on the interferometer in the opposite direction to its motion. Explicitly, Lorentz \cite{lorentz,lorentz1,lorentz0} assumed that if one arm were longer than the other by the quantity $\frac{1}{2}l_0\beta^2$ the result of the experiment would be zero. He conjectured that the dimension parallel to the direction of motion were changed in the proportion of 1 to $1+\kappa$ and that perpendicular in the proportion of 1 to  $1+\epsilon$, hence, the next equation must be true
\begin{equation}
\label{loreq}
\epsilon-\kappa=\frac{1}{2}\beta^2.
\end{equation}
From here it might be that
\begin{itemize}
\item [(i)] $\epsilon=0$, $\kappa =-\frac{1}{2}\beta^2$,
\item [(ii)] $\epsilon=\frac{1}{2}\beta^2$, $\kappa =0$,
\item [(iii)] $\epsilon=\frac{1}{4}\beta^2$, $\kappa =-\frac{1}{4}\beta^2$.
\end{itemize}
In fact, Lorentz oversimplified the number of solutions because, strictly speaking, there could exist an infinite number of them. However, based on previous works of Oliver Heaviside \cite{heaviside}, who had shown in 1888 that the corresponding components of the electric field $\mathbf{E}$ of a charge in motion altered by a factor $\gamma^{3}$, and on their own calculations, Lorentz \cite{lorentz2,lorentz1,lorentz4}, FitzGerald \cite{fitzgerald,sorensen}, and Larmor \cite{larmor1,larmor2} determined that objects must have shorten by the factor $\gamma^{-1}$ to keep the state of equilibrium in the direction of motion. Thus, the most plausible option is (i).

So far, guided by the experiment we have resolved that the longitudinal arm shortens by the factor $\gamma^{-1}$ while the perpendicular arm remains unaltered. Consequently, replacement of $l_0$ by $l_0\gamma^{-1}$ in the corresponding equations yields
\begin{equation}
\label{tnew}
t_{\|}=t_{\bot}=\frac{2l_0}{c}\gamma,
\end{equation}
which explains why there is no fringe shift. 

Also due to length contraction of the HSM we must incorporate in the formula \eqref{d1} the factor $\gamma^{-1}$ so that the angle of reflection \eqref{efangle}  is now given by
\begin{equation}
\label{neeffe}
\tan \frac{\alpha_\pm}{2}=\pm \frac{\sqrt{1\pm\beta}-\sqrt{1\mp\beta}\tan \varphi_0}{\sqrt{1\pm\beta}+\sqrt{1\mp\beta}\tan \varphi_0}.
\end{equation}

\subsection{Time dilation}
\label{timedil}
Next, if we identify the quantity $2l_0/c$ as the time $t_{\bot 0}=t_{\| 0}$ that light takes to travel the distance $2l_0$ at rest in $S$; a simple inspection of Eq. \eqref{tnew} reveals that this same process is delayed for the observer in $S'$ by the factor $\gamma$. This realization constitute a strong evidence that because of the aether is not dragged along with the moving source, the processes of electromagnetic nature are delayed for an observer in motion. As a consequence every period $T$ of the frequency must be also delayed by the same factor, and the frequencies of both frames will be related as $\nu'=\nu \gamma$.

\subsection{The two-way speed of light}
\label{twowaysp}
Now in approach B we were expecting that $\delta s'=const.=0$ and $\delta V'(v) \neq const.$ If item (\textbf{\dag}) is true, it demands not only that \emph{the velocity of the waves must be the same in all directions but also that Galilean transformations are flawed}. 

To determine the value of the speed of light waves in $S'$ we must consider the conclusions from the observer in $S$. Accordingly, the earth observer has to replace in the formula \eqref{t2}, $l_0$ by $l_0\gamma^{-1}$ and, $t_i'$ by $t'_i\gamma$. And for the transversal direction in Eq. \eqref{abc}, $t'_{\bot}$ by $t'_{\bot} \gamma$. On doing this, he would arrive at 
\begin{equation}
\label{abcsd}
t'_{\bot}=t'_{\|}=\frac{2l_0}{c}. 
\end{equation}
Evidently, this expression not only resembles the calculations performed at rest in $S$, but also shows that $\delta V'=0$ which explains that in $S'$ there is no fringe shift as well. With these results we have arguments to surmise that a two-way measurement of the speed of light in any other inertial frame different from $S$ and in any direction must be $c$. It comes about that the observer in $S'$ must resolve that a wave equation of the same form of the Eq. \eqref{waveeq} must govern electromagnetic wave phenomena. 

It is important to remark that in Eq. \eqref{abcsd} we have only determined the \emph{average or two-way speed of light} for the longitudinal and transversal directions, but this equation does not give us information about the speed of light for each of the four paths. Recall that there is still aether wind and we expect, at least, a speed difference for the backward and forward advances in the longitudinal direction. A calculation for the one-way speed of light found by me \cite{perez} and other authors \cite{mansouri,guerra2,abreu,iyer1} has shown that, for symmetry reasons, the transversal speed of light is $c$, but for the longitudinal directions the one-way speeds are given by $c'_{\pm}=c(1\pm \beta)^{-1}$, respectively. Despite this apparently contradictory result, it can be easily shown that \emph{the two-way speed of light in any direction is always} $c$ \emph{in agreement with SR} (see section \ref{tangher} below). Hence, insertion of these values into Eq. \eqref{nfringe3} proves also that $\delta V'=0$ and therefore $\delta'=0$, so there is no fringe shift.

\subsection{Preliminary conclusions}

In summary, both observers have arrived at the following conclusions:
\begin{itemize}
 \item [(a)] The laws of electrodynamics are at variance with Galilean transformations.
  \item [(b)] The dimension of the arm in the direction of motion shortens by the factor $\gamma^{-1}$.
  \item [(c)] The dimension of the arm in the transversal direction is not affected.
  \item [(d)] Electromagnetic processes are delayed by the factor $\gamma$ for inertial frames different from the aether.
  \item [(e)] When measured in vacuum and in inertial frames the two-way speed of electromagnetic waves must be isotropic for all inertial observers, and thus a universal constant.
\end{itemize}

\subsection{Understanding and solving the problem}
Notice that, so far, our preliminary conclusions are based only on the experimental facts. From item (e) one can infer that the same wave equation governs electromagnetic phenomena in all frames (the MM experiment suggest this). Indeed, we have followed the line of thought of Lorentz, FitzGerald and Poincar\'e, the next step is to find a general theory that radically explain the phenomena. Also, it would be pertinent to point out that we have essentially arrived at the two postulates of SR, which, obviously, they did not know it. Thus, the natural step to be taken was to search for a new set of transformations that replaced \eqref{galtrans} and that harmonized not only with these results but also with classical mechanics \cite{granek}. Actually, this is the job that Larmor \cite{larmor1,larmor2}, Lorentz \cite{lorentz2,lorentz0,lorentz4}, Poincar\'e \cite{poincare2} and Einstein \cite{lorentz0,einstein2} did when they discovered the Lorentz transformations:
\begin{equation} 
\label{lortrans}
x'  =  \gamma(x-vt), \qquad t' = \gamma(t-vx/c^2).
\end{equation}

Worried about the Galilean principle, Einstein had a great insight and realized that the law of the constancy of the speed of light contained in Maxwell's electrodynamics was at variance with the theorem of addition of velocities. He then determined to postulate that, just in the same way that the laws of mechanics hold for all inertial observers, the Maxwell's equations must hold good too. In addition, he also postulated the constancy of the speed of light for all inertial observers. From these two postulates and the definition of simultaneity the Lorentz transformations were derived. In fact, a closer look of the second postulate shows that it appears as a tautology of the first one. By demanding the invariance of the physical laws for all inertial observers, Maxwell's equations must be invariant and therefore the speed of light in vacuum too.

Favoring the work of Einstein, Minkowski \cite{minkowski} just did the mapping from the physical to the geometrical world through the introduction of a new metric, namely: $ds^2=(cdt)^2-(dx^2+dy^2+dz^2)=(cdt')^2-(dx'^2+dy'^2+dz'^2)$ in which the two postulates are also implied. This metric generalizes the traditional Euclidean space where the Galilean group resides. And although the geometrization of SR was a great success for theoretical physics; from the perspective of phenomenological physics, such mathematical interpretation overshadows even more not only the atomic mechanisms of time dilation and length contraction, but also the electromagnetic medium. This is due, from the theoretical standpoint, to the erroneous belief that there is no privileged frame. Einstein was convinced that any inertial frame was equivalent for the description of physical phenomena and, in this sense, he was right; from this perspective the notion of a privileged frame in the old aether theory (OET) becomes superfluous. But I deeper reflexion shows us that this is not the case, the rejection of the aether is equivalent, from the view posed here, to the rejection of space. Secondly, from the experimental standpoint, Einstein trusted \footnote{{\scriptsize It has been argued that Einstein did not know about the MM experiment, but there is a clear evidence in section 9 (Transformation of the Maxwell-Hertz equations when convection...) of Einstein's 1905 paper when he said: \emph{If we imagine the electric charges to be invariably couple to small rigid bodies (ions, electrons), these equations are \textbf{the electromagnetic basis of the Lorentzian electrodynamics and optics of moving bodies.} ...we have the proof that, on the basis of our kinematical principles, \textbf{the electrodynamic foundation of Lorentz's theory of the electrodynamics of moving bodies} is in agreement with the principle of relativity.} Obviously, he was referring to the Lorentz's paper of 1904, where the Michelson-Morley experiment is treated \cite{lorentz2}.}} the spurious conclusions drawn from the MM experiment \cite{einstein2,einstein10}. As we have shown above the experiment only tests length contraction and time dilation for inertial frames in motion relative to $S$;  which at the end leads us to conclude that the phase remains constant because the speed $v$ is canceled out. Now we show this for the Kennedy-Thorndike experiment.

\subsubsection{Nullifying effect in vacuum experiments: the Kennedy-Thorndike experiment}
\label{kenthorn}

The Kennedy-Thorndike experiment \cite{kennedy} is a modified version of the MM experiment in which the arms of the interferometer are of unequal lengths. Hence, in the equations of approach A we just have to replace $l_0$ by $l_{\bot}=l'_{\bot}$ and $L_{\|}$, respectively. Based on either SR or our preliminary conclusions, i.e., Lorentz-FitzGerald contraction $L_{\|}=l'_{\|}\gamma^{-1}$, we would find that, as seen from $S$, the time shift is
\begin{equation}
\label{dtmin}
\delta t=2\frac{\delta l'}{c}\gamma.
\end{equation} 
where $\delta l'=l'_{\|}-l'_{\bot}=$constant. But since the experiment is performed in $S'$, recall from section \ref{fringeshift} that the fringe shift is also frequency dependent, cf. \eqref{fringe2}. So, we must take into account our previous findings that $\omega'=\omega \gamma$, therefore
\begin{equation}
\label{dtmin}
\delta' =2\omega' \frac{\delta l'}{c}.
\end{equation}
Since $\omega'=$constant the phase difference is also constant, and by the fringe shift conditions there is no induced fringe shift.

It is fair to show the path that Kennedy and Thorndike followed to solve this problem. Let us see what they did. First, they based their work on the following assumptions:
\begin{itemize}
  \item [(x)] \textsf{There exists at least one coordinate system in which Huyghens principle is valid and the velocity of light is the same in all directions. This assumption is unobjectionable from the standpoint either of relativity or of any plausible hypothesis involving an ether; for relativity, it is true for all uniformily moving systems, and in the latter case for any system at rest in the ether.}
  \item [(xx)] \textsf{The MM experiment indicates that a system moving with uniform velocity $v$ with respect to such a system has dimensions in the direction of motion contracted in the ratio $\gamma^{-1}$ as compared to dimensions in a fixed system, while dimensions perpendicular to this direction are unchanged. This is in part assumption, for although there can be little doubt that the experiment yields a strictly null result, nevertheless it actually shows only that dimensions in the direction of and perpendicular to the motion are in the ratio mentioned; either of these dimensions might be any function of the velocity so long as the ratio is preserved.}
\end{itemize}

Note that assumption (x) is similar to statement (a) given above. Later they proceeded to perform the calculations as seen from the frame $S$, in which they explicitly incorporated Lorentz contraction (from SR) in the longitudinal direction. As a consequence they arrived at the time difference
\begin{equation}
\label{dtken}
\delta t'=\frac{\delta s'}{c}\gamma,
\end{equation}
where $\delta s'=s'_2-s'_1$ is the path difference and $s'_i$ ($i=1,2$) is the total distance traveled by each wave as measured in $S'$. Since the arm lengths are fixed, $\delta s'$ is a constant. Hence, the number of fringes is given by
\begin{equation}
\label{difn}
N=\nu \frac{\delta s'}{c}\gamma,
\end{equation}
with $\nu$ the frequency of light as measured in $S$ (cf. \eqref{fringe1}). 

Later they discuss that: 

\textsf{The quantity $\nu$ is the only one whose possible variability with velocity remains considered. From the standpoint of relativity, $\nu=\nu'\gamma^{-1}$ where $\nu'$ is the constant value of the frequency which would be determined with standards moving with $S'$; this value of $\nu$ would evidently make $N$ constant. If, on the other hand, $\nu \neq \nu' \gamma^{-1}$ these (Lorentz) transformations do not apply and it turns out that there exists but one system $S$ satisfying assumption (x); this unique system would be the absolute reference frame postulated in the classical ether theory. In this case $N$ is evidently a function of the velocity of $S'$ with respect to the absolute reference frame. Evidently, then, the relativity hypothesis can be tested by determining whether $N$ is a constant as $v$ changes in consequence of the motions of rotation and revolution of the earth.}

Let us paraphrase this reasoning: Kennedy and Thorndike thought that the aether was part of a theory governed by GR. And because of, in the MM experiment, the arms are of equal length, they thought that such experiment was only sensitive to length contraction. Also, we can assert that, under GR, they arrived at the conclusions that both $\nu \neq \nu'\gamma^{-1}$ and the one-way speed of light in other frames is not isotropic. So, in essence, they followed approaches A and B above. But, in this case, they reasoned in the opposite direction. In SR the aether does not exist, the one-way speed of light is isotropic in all frames and the expression $\nu'=\nu\gamma^{-1}$ is fulfilled. If they found a fringe shift that do not match the predictions of SR, that would imply that SR is incorrect and that the aether does exist. So, here we see that the problem arises from the initial assumptions. They are still observing as one phenomenon the Galilean transformations (the mathematics) and the concept of the aether (the baggage \footnote{To deeply feel the meaning of the word ``baggage", see the work of Max Tegmark \cite{tegmark5}.}).

It is unnecessary to go further into the details of the experiment. It is only important to mention that the experiment was realized in vacuum conditions and that the results were almost zero; so that the same conclusions to those of the MM experiment were reached. However, these results do not constitute a strong evidence against the aether. They only expose actual length contraction and time dilation (see test theories of SR \cite{mansouri,robertson,macarthur0,zhang}). Moreover, we have determined that, in $S'$, the experiments also require that the two-way speed of light be a constant. 

Now let us reexamine the experiment in non-vacuum conditions and reason whether the aether can be detected or not.
 
\subsubsection{Non-vacuum experiments}
\label{novac}

Michelson, Morley, Miller et al. perfectly knew that their results were never zero \cite{michelson0,michelson1,miller,miller2}, but it was not quite clear for the physicists of that time whether the data was showing speed anisotropy from the Galilean prediction or any other effect, e.g., gravitational. Now that we have assimilated the old and the new theories, we can revisit the facts. Next we explain the basic idea behind this. Of course, angular and gravitational effects can be introduced, but since we are interested in the physics and not in detailed calculations, we shall restrict ourselves to show only that when a rarified gas is introduced the velocity $v$ is not canceled out from the equations. 

For convenience the experiment will be judged from the frame $S$. Let us suppose that we fill the arms of the MM interferometer with a rarified gas with refractive index $n_g$ measured in the rest frame. Then the speed of the waves as measured in $S$ will be given by $V_i=c/n_g$, with $i=1,..,4$. If the interferometer is at rest in $S$, we have $\delta V=0$ and $\delta s=0$, therefore there is no fringe shift. But when the interferometer is set in motion we must follow a similar procedure as in approach A. In such a case, the total distance traveled by the transversal waves is
\begin{equation}
\label{haz1g}
s_{\bot}\equiv Vt_{\bot}=\sqrt{(vt_{\bot})^2+(2l')^2},
\end{equation}
with $l'$ the arm lengths measured at rest. On solving for the time yields
\begin{equation}
\label{ttransg}
t_{\bot}=\frac{2l'}{V}\eta,
\end{equation}
where $\eta=1/\sqrt{1-(n_g\beta)^2}$. For the longitudinal direction the wave fronts travel the distances
\begin{equation}
\label{s1g}
s_{1} \equiv Vt_1= l+vt_1, \qquad s_{2}\equiv Vt_2= l-vt_2,
\end{equation}
for the forward and backward advance, respectively; where $l=l'\gamma^{-1}$. Solving for the times from the previous equations we find that
\begin{equation}
\label{t1g}
t_1=\frac{l}{V-v}; \qquad t_2=\frac{l}{V+v}.
\end{equation}
So, considering that $t_\|=t_1+t_2$, we obtain
\begin{equation}
\label{tparalg}
t_\|=\frac{2l'}{V}\eta^2\gamma^{-1}.
\end{equation}
Recall that $\delta=\omega \delta t=\omega' \gamma^{-1} \delta t$. Thus, subtracting \eqref{ttransg} from \eqref{tparalg} and expanding in powers of $\beta$, the fringe shift is given by
\begin{equation}
\label{deltatg}
\delta' \approx \omega' l' \frac{(n_g^2-1)}{V}\beta^2+O(\beta^4).
\end{equation}
This equation clearly shows that in spite of the introduction of the frequency relation the fringe shift is still function of the earth speed. But here again we face the same problem that we found at the end of approach A. As long as we consider the earth as an inertial frame, $\delta'=$const. hence no fringe shift is induced and therefore the principle of relativity still holds. 

\subsubsection{Experimental proofs}

\begin{figure}[htp]
\begin{center}
\includegraphics[width=9cm]{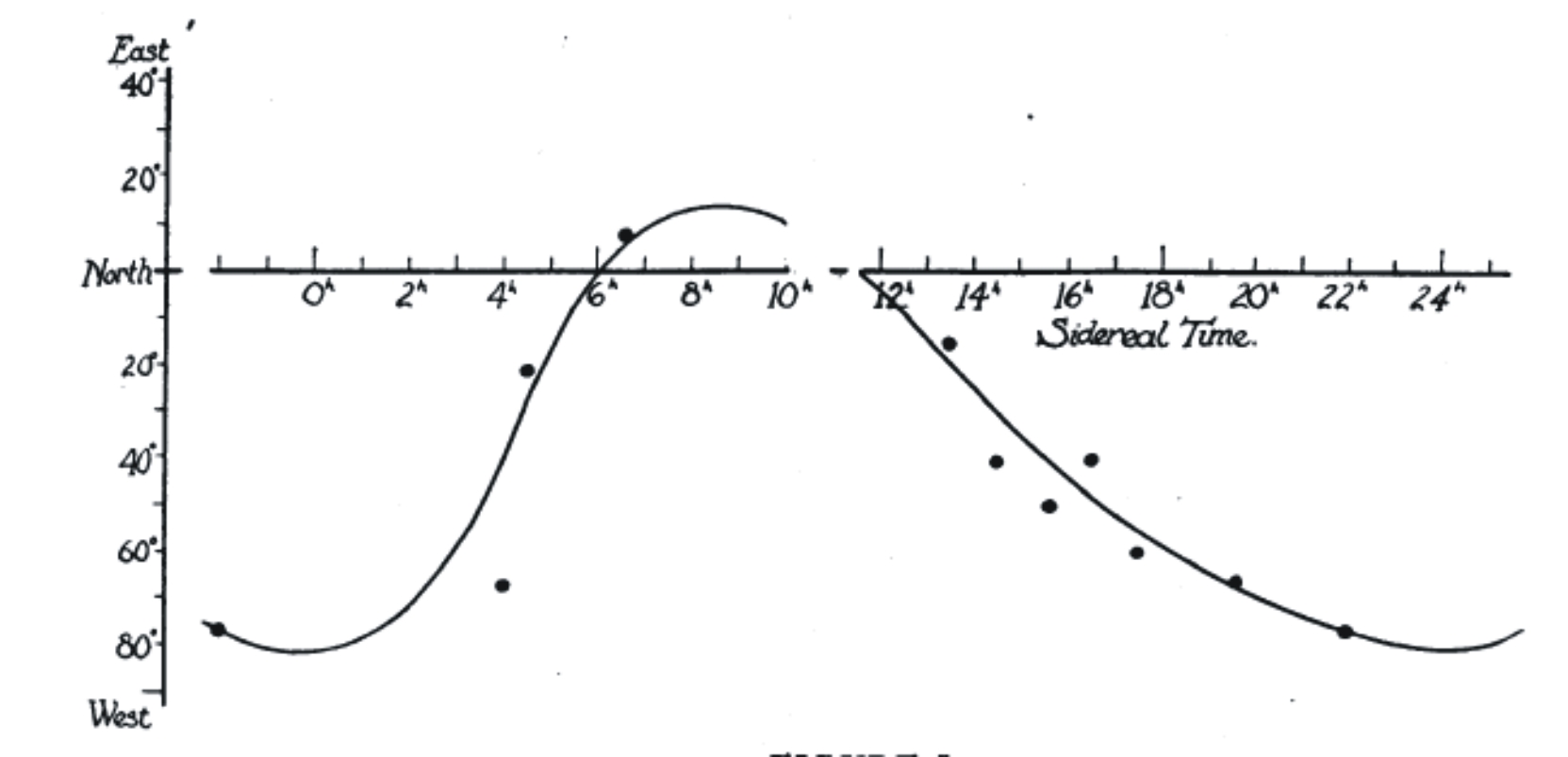}
\caption{Results of the Miller experiment \cite{miller,miller2} carried out in 1925. The curve has been drawn just to guide the eye.}
\label{millerf}
\end{center}
\end{figure}

The result obtained in Eq. \eqref{deltatg} at least gives us hope to seek for the preferred frame using interferometric experiments realized in non-vacuum conditions. What is important to mention in this section is that the fringe shift was not null. In Fig. \ref{michelson} we have shown the results obtained by Michelson and Morley in the experiment of 1887 and in Fig. \ref{millerf} we present the curve for the experiment of Miller \cite{miller,miller2} conducted in 1925.
Today most physicists, upholders of the vision of non-privileged frames, refer to the work of R. Shankland \cite{shankland,shankland1} and other vacuum experiments \cite{jaseja,brillet,riis,hils,muller,muller1,chen,herrmann,antonini,wolf} to witness that local Lorentz invariance has not yet been violated and that the postulates of SR still stand up. Shankland disapproved the work of Miller alleging statistical fluctuations on the readings of the fringe positions and temperature fluctuations. However, H. Munera has carefully reanalyzed the experimental data of the MM, Miller and other experiments. He has discovered a number of systematic errors that have been overlooked in the literature and shows that the results of the experiments are not null
\cite{munera,munera1}. When these effects are considered the curves of Figs. \ref{michelson} and \ref{millerf} are significantly enhanced. More recently researchers are not only reanalyzing the old and new data but also planning more accurate aether-drift experiments under this new vision \cite{guerra2,abreu,marinov1,cahill,consoli,consoli1,consoli2,guerra3,spavieri,consoli3,lammerzahl1,lammerzahl2}.

At this point, we must emphasize that our fringe shift conditions demand that in order to observe a fringe shift something must be varying in time, in this case $v$, therefore the non-null result rather suggests that SR is only an adequate description for inertial frames and a satisfactory explanation of the problem can only be found in a more general theory. The extension of the physics to non-inertial frames was also worked out by Einstein in 1911 in the development of the principle of equivalence \cite{einstein4}. Later Hilbert gave the complete solution to the problem \cite{hilbert} and few days after Hilbert's publication, Einstein \cite{einstein5} published his version with the introduction of the general theory of relativity (GTR). Unfortunatelly, SR and GTR are theories in which the aether does not enter and thus they lack the notion of a privileged frame and of space as a material continuum. Under such scenario, those who believe in these matters must look for other alternatives. Fortunately, such theories are already developed although not widely known. Let us discuss the first approach analogous to SR but with a privileged frame.

\section{A new \ae ther theory}
What we have seen, so far, is only a survey, on the one hand, to identify the key points of the problem. We have shown how the situation can be kinematically and partially resolved from the standpoint of two philosophies: FitzGerald-Larmor-Lorentz-Poincar\'e (with aether) and SR (without aether). On the other, we have explained that, for experiments in vacuum conditions, the consideration of relativistic effects in our calculations makes the speed of the earth to vanish, but the situation acquires a novel aspect when a rarified gas is introduced. 

For the physicist educated under the paradigm of modern theories, the arguments given by Lorentz about relativistic effects seems so artificial and of null generality. To satisfy this inquietude below we put forward a new alternative for an aether theory which makes clear some of the puzzles left by SR, e.g., the actuality or appearance of relativistic effects.

\subsection{Actual or apparent?}
\label{acap}
In these respects, I consider that the present work helps to elucidate the perplexity kept in the mind of several physicists about the actuality of relativistic effects. Frequently, it is found in the literature \cite{shankland1,holton0,shankland0,field} dismissals of the FitzGerald-Lorentz contraction which are commonly referred as ``\emph{ad hoc} assumptions", or ``auxiliary hypotheses". Nevertheless, we have readily shown that when the MM experiment is judged from the perspective of the observer in $S$, $\delta$ is only function of the PPL whose limits are only constrained by the arms of the interferometer, therefore such \emph{assumptions} become the most physical and logical conclusions that can be drawn because the PPL are the actual distances traversed by the light waves in $S$. It is worth to point out that there is a marked difference between the \emph{conclusion drawn} from the condition $\delta s=0$ and the \emph{auxiliary hypotheses} that Lorentz had to conceive to explain the shortening of the arms. Physicists should understand that, if we believe in our condition, the object contraction is actual and, certainly, the assumptions to solve the problem of object contraction may, perhaps, sound for some rather artificial; Lorentz himself was aware of this point \cite{lorentz2}.

On the other hand, when Lorentz discovered his transformations \cite{lorentz2,lorentz0,lorentz4} he correctly called the time $t'$ the \emph{local time} (proper time in modern terms) in order to distinguish it from the \emph{universal time} $t$ (or absolute time) that belongs to the system $S$. Following Einstein's definition of clock synchronization \cite{einstein2} we may place a network of clocks all over the system $S$. Since all inertial systems ($S'$, $S''$, etc.) are logically either at rest or in motion relative to $S$, the local time $t'$ does not interfere with the flow of the universal time $t$. However, if we reject the existence of $S$ then we are left only with an infinite set of inertial systems from which only relative motion, local times and relative spaces can be spoken of. And in this sense the ``local times" and lengths of SR acquire the same prominence for all observers. The cost to be paid for such refusal is a series of paradoxes, e.g., the Supplee and twin paradoxes \cite{frisch,minguzzi,jones,west,dewan,rindler,nawkocki,dewan1,nicolic,pierce,supplee,matsas}, which cannot be solved without the assistance of the GTR. 

\subsection{The Tangherlini transformations}
\label{tangher}
To be free from paradoxes and avoid the GTR we must accept a privileged frame and support the opinion that relativistic effects are actual. The weighty arguments for a new aether theory (NET) were given by Mansouri and Sexl \cite{mansouri,iyer1,tangherlini,marinov,petry}. They found the so-called Tangherlini transformation
\begin{equation}
\label{tmans}
t'=t\gamma^{-1}, \qquad x'=\gamma(x-vt),
\end{equation}
that only differs from the Lorentz transformation by the convention in the clock synchronization. They showed that \emph{a theory that retains absolute simultaneity is equivalent to SR}.  

Researchers \cite{guerra2,abreu,iyer1,guerra3,spavieri,marinov,petry,selleri,levy,puccini} have studied in more detail the properties of these transformations which, evidently, have no symmetrical inverse:
\begin{equation}
\label{itmans}
t=t'\gamma, \qquad x=\gamma^{-1}x'+\gamma vt'.
\end{equation}
They have pointed out that one of the consequences of this asymmetry is that the position of the origin of $S$, $x = 0$, is given in $S'$  by $x'=-\gamma^2vt'$ which implies that the speed of $S$ relative to $S'$ is $v'=-\gamma^2v$, and not just $-v$ as one could expect from the OET or SR. Since objects really contract and clocks really run slower in $S'$, one factor $\gamma$ accounts for length contraction and the other for time dilation. Notice also that for $v\ll c$ the above transformations reduce to the Galilean limit.

As an illustration of the value of this theory, it is worthy to consider the case in which we apply to the wave equation \eqref{waveeq} the transformations \eqref{tmans}. After a straightforward calculation the equation in $S'$ acquires the perspicuous form:
\begin{equation}
\label{netwaveq}
\biggl[\frac{\partial^2}{\partial x'^2}+\frac{2v}{c^2}\frac{\partial^2}{\partial x' \partial t'}-\frac{\gamma^{-2}}{c^2}\frac{\partial^2}{\partial t'^2}\biggr] \Psi'=0.
\end{equation}
The reader can easily verify that the application of the inverse transformations \eqref{itmans} to this equation recovers the original wave equation. Here we have just found one of the advantages of the NET over the OET. To evidence this, the wave equation \eqref{waveeqin} in one dimension is
\begin{equation}
\label{waveqnet}
\biggl[\gamma^{-2}\frac{\partial^2}{\partial x'^2}-\frac{2v}{c^2}\frac{\partial^2}{\partial x' \partial t'}-\frac{1}{c^2}\frac{\partial^2}{\partial t'^2}\biggr] \Psi'=0.
\end{equation}
However, the inverse Galilean transformations do not restore this equation to that in the aether frame. In fact, no kinematic transformation of $\Psi'$ can restore it \cite{jackson}. This fact can be considered as another clear evidence of the incompatibility between Maxwell's formulation of electrodynamics and the Galilean transformations.

Now we obtain a couple of important results to show that the NET gives the same kinematical information as that of the SR. We have seen in section \ref{fringeshift} that the phase of a wave give us information on how to obtain the number of fringes. If this phenomenon is observed in one frame, it must be the same in any other frame since it is just a counting procedure. Thus, the phase of wave must be an \emph{invariant quantity}. To avoid the complications of solving the wave equation \eqref{netwaveq} we rather make use of this principle. Hence in one dimension we have
\begin{equation}
\label{varinv}
\Phi=\omega\biggl(t-\frac{x}{c}\biggr)=\omega'\biggl(t'-\frac{x'}{c'}\biggr),
\end{equation}
where the prime variables are related to the frame $S'$. Applying the transformations \eqref{tmans} to this expression, we obtain after a little algebra the following relations
\begin{eqnarray}
\label{csprima} c'_{\pm} & = & c(1\pm\beta)^{-1}, \\
\label{comega} \omega'_{\pm} & = & \omega \gamma(1\pm \beta),
\end{eqnarray}
where the signs $\pm$ stand for the forward and backward advances, respectively. Since the two-way  speed of light is the one that is \emph{experimentally} measured \cite{abreu,spavieri}, what we measure is, actually, the harmonic mean \cite{perez,iyer1}, that is:
\begin{equation}
\label{cmean}
\bar{c}=\frac{2}{\frac{1}{c'_+}+\frac{1}{c'_-}}=\frac{2}{\frac{1}{c(1+\beta)^{-1}}+\frac{1}{c(1-\beta)^{-1}}}=c.
\end{equation}
Expressions \eqref{comega} and \eqref{cmean} show the agreement between the NET and SR. Iyer and Prabhu \cite{iyer1} have also theoretically worked out the one-way speed of light in $S'$, they found the more general relation:
\begin{equation}
\label{oneway}
c'=c\biggl(1\pm \frac{\mathbf{\hat{n}'}\cdot \mathbf{v}}{c}\biggr)^{-1}, 
\end{equation}
where $ \mathbf{\hat{n}'}$ is a unit vector that points in the direction of energy flow of the light beam as determined in $S'$ and $\mathbf{v}$ is the velocity vector of $S'$ relative to $S$. Following our example the reader can easily verify from this equation that the harmonic mean is $c$, in other words, that the two-way speed of light is a universal constant for all inertial frames in agreement with SR and experience. Another useful result is the theorem of addition of velocities. From the Tangherlini transformations we directly obtain
\begin{equation}
\label{addthe}
u'=\gamma^2(u-v),
\end{equation}
where $u'$ is the one-way speed of the particle as seen in $S'$ and $u$ is the one-way speed of the particle as seen in $S$. With $u=\pm c$ we obtain again Eq. \eqref{csprima}.

\subsection{Light behavior within the context of the NET}

\begin{figure}[htp]
\begin{center}
\includegraphics[width=4.5cm]{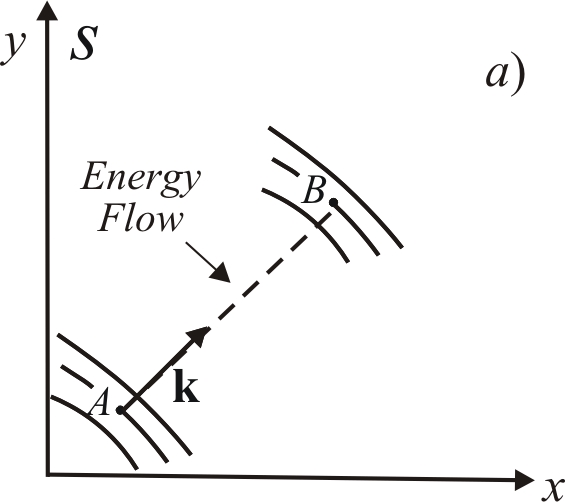} \hspace{1cm}\includegraphics[width=5cm]{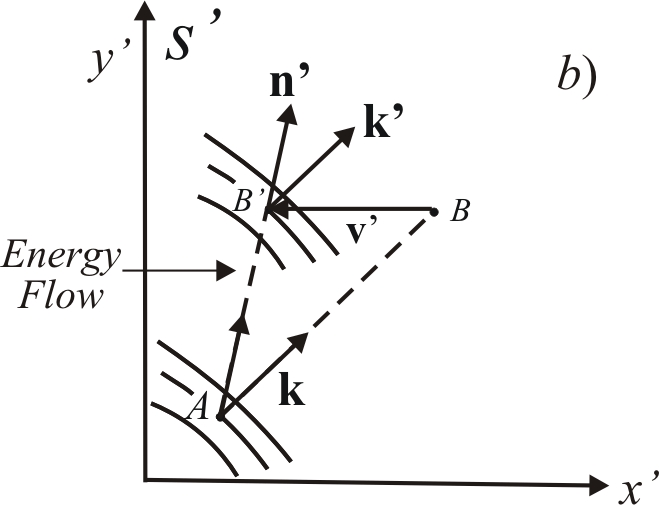}
\caption{Relation between the unit vectors $\mathbf{\hat{n}'}$ and $\mathbf{\hat{k}}$. ($a$) In $S$, both the wave normal and the energy flow point in the same direction $\mathbf{\hat{k}}$. ($b$) In $S'$, the wave normal still points in the direction $\mathbf{\hat{k}'}=\mathbf{\hat{k}}$ but, due to the aether wind, the energy flow points in the direction $\mathbf{\hat{n}'}$.}
\label{unitvec}
\end{center}
\end{figure}

Let us make clear the behavior of light among two inertial observers. First of all, it is useful to express the unit vector $\mathbf{\hat{n}'}$, from Eq. \eqref{oneway}, in terms of quantities only measured in $S$. For this purpose consider the wave fronts depicted in Fig. \ref{unitvec}. These segments can be seen as wave packets. Also assume that the frame $S'$ moves relative to $S$ at speed $\mathbf{v}$.  At $t=t'=0$ the center of the wave fronts are at point $A$ as seen in both frames $S$ and $S'$. 
In $S$ the wave fronts have the direction $\mathbf{\hat{k}}$ and they will arrive at point $B$ after a unit time $t_u$ traversing a distance $\overline{AB}=ct_u$. However, as seen from $S'$, the wave packet arrives at point $B'$ after a unit of time $t'_u=\gamma^{-1}t_u$ traversing a distance $\overline{AB'}=c't'_u=\gamma^{-1}t_uc^2(c+\mathbf{\hat{n}'}\cdot \mathbf{v})^{-1}$. Because of the Tangherlini transformations this point differs from point $B$ by the vector $\mathbf{v}'=-\gamma^2\mathbf{v}$, as shown in Fig. \ref{unitvec} (b). In $S'$ the direction of the energy flow of waves is not parallel to $\mathbf{\hat{k}}$ but along the unit vector $\mathbf{\hat{n}'}$ which is given by
\begin{equation}
\label{kkk}
\mathbf{\hat{n}'}=\frac{c\mathbf{\hat{k}}-\gamma^2\mathbf{v}}{|c\mathbf{\hat{k}}-\gamma^2\mathbf{v}|}.
\end{equation}
This equation emphasizes the crucial role played by the effective angle found in section \ref{misang}. Such effect is commonly overlooked in most kinematical experiments where waves are involved and therefore, if omitted from consideration, it gives a false idea of light propagation. 

Bearing in mind the MM experiment and the angular effect, we have seen that, from $S$, the direction of both the energy flow and the wave normal in the forward-transversal direction are diagonal and given by Eq. \eqref{neeffe} with vector $\mathbf{\hat{k}_{\bot}}$ (Fig. \ref{unitvec} ($a$) resembles this situation). But due to the aether wind and using Eq. \eqref{kkk}, the observer in $S'$ determines that the energy flow points along the unit vector $\mathbf{\hat{n}'}_\bot$; in spite of this, the wave normal still points in the direction $\mathbf{\hat{k}'_{\bot}}=\mathbf{\hat{k}_{\bot}}$ (Fig. \ref{unitvec} ($b$) corresponds to this situation). If there were no aether, the direction of energy flow in $S'$, after interaction with the HSM, would be in the direction of $\mathbf{\hat{k}_{\bot}}$, that is, the observer would see the beam trajectory following a diagonal in contradiction with experience.

Now that we are dealing with the propagation of light, let us qualitatively discuss another important source for the fringe shift that we have not regarded in the first sections to avoid confusion. In Fig. \ref{fisk} we have shown the plane of the apparatus placed parallel to the motion. When we performed the calculations for the fringe shift in approach A we only considered a factor of 2 to compensate the $\pi/2$-rotation of the interferometer, but we did not discuss the behavior of the wave fronts when they are emitted perpendicularly to the direction of motion and we also dismissed the mirror effects in such situation. 
\begin{figure}[htp]
\begin{center}
\includegraphics[width=5cm]{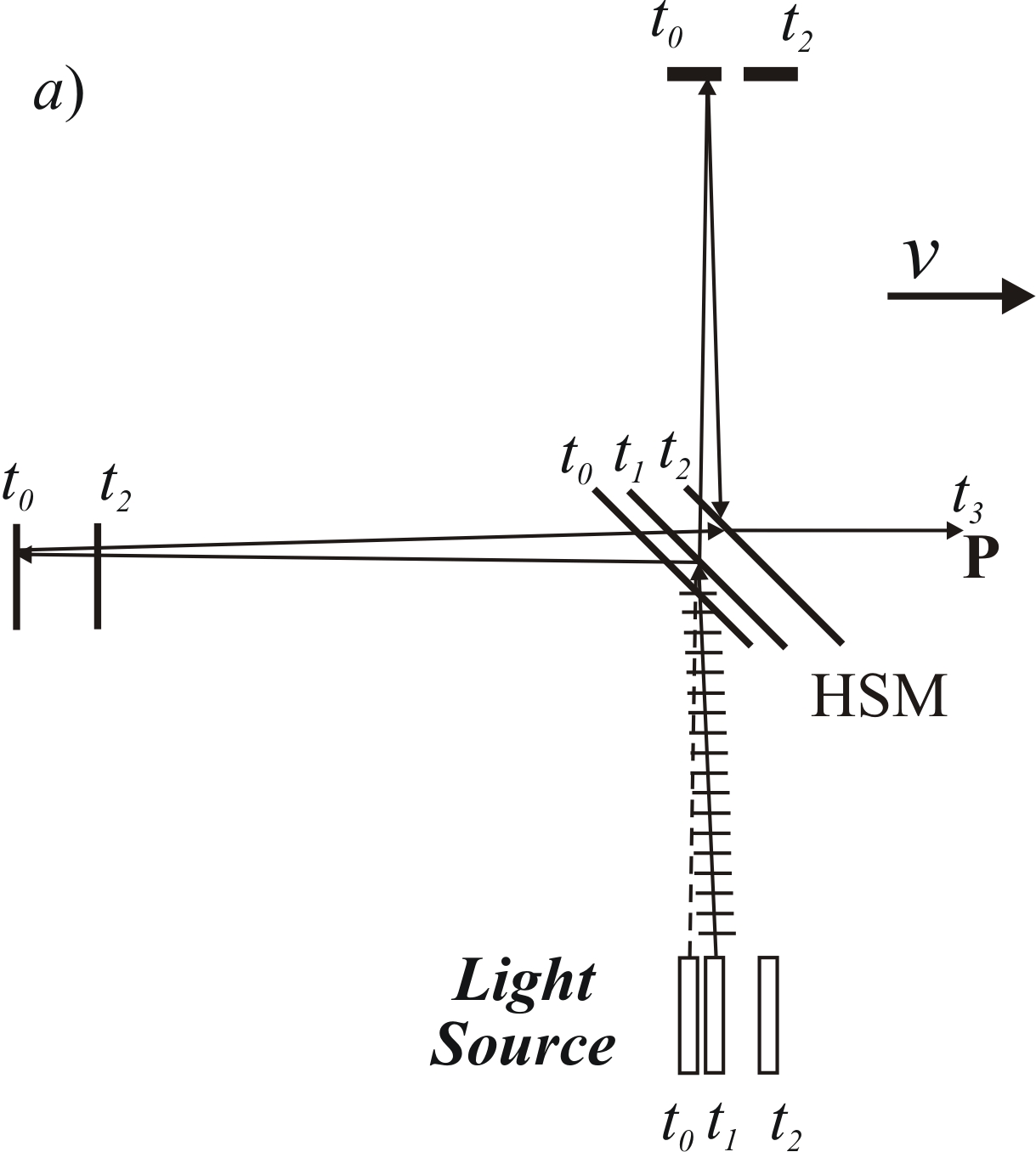} \hspace{2cm}\includegraphics[width=3.2cm]{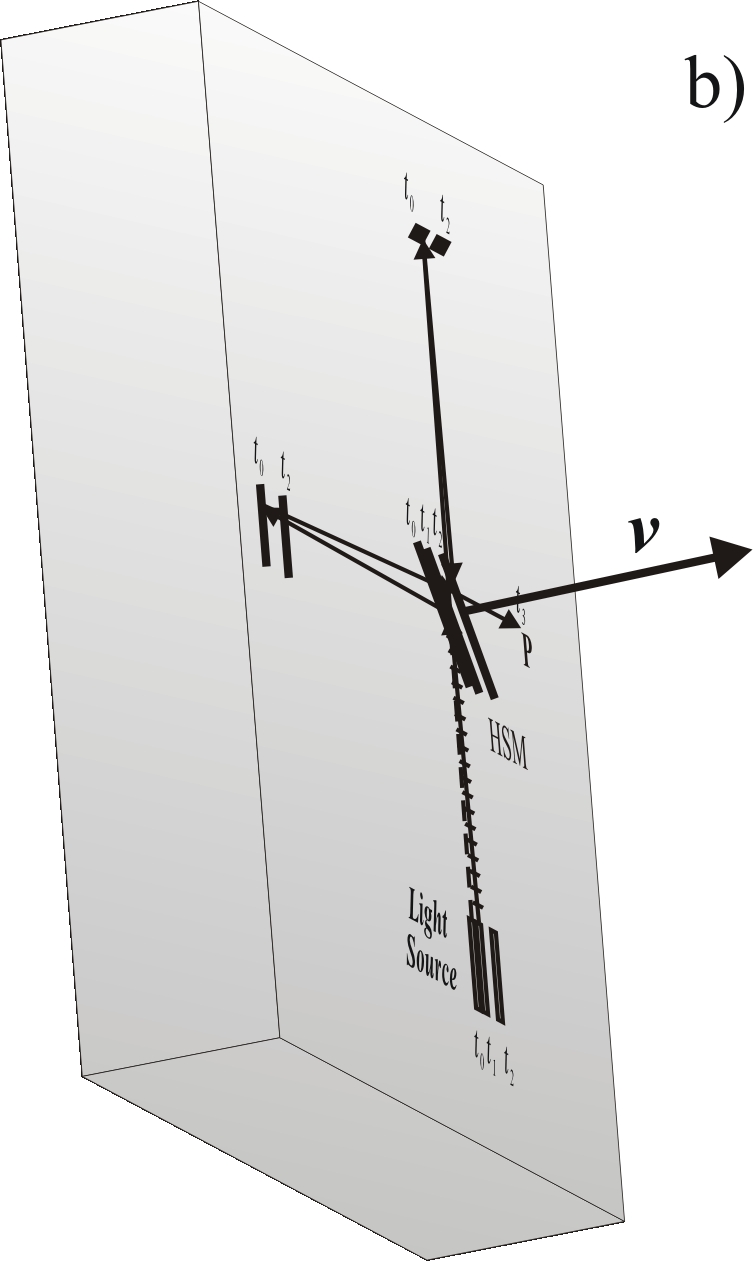}
\caption{The interferometer as seen from the aether frame. ($a$) The plane of the interferometer is parallel to the direction of motion and the apparatus has been rotated $\theta=\pi/2$ rad. A variation of the phase $\delta$ may be expected when the angle $\theta$ varies. Thin arrows represent the vectors of the energy flow. (b) The plane of the interferometer is orthogonal to the motion. Due to this the interferometer does not suffer length contraction and, provided that we considered the wave reflection at the mirrors, the problem may be reduced to the case at rest in $S$. In such situation, even when $\theta$ changes, $\delta$ remains constant and no fringe shift is expected.}
\label{introt}
\end{center}
\end{figure} 

The interferometer rotated an angle $\theta=\pi/2$ rad is shown in Fig. \ref{introt}($a$). Here we can distinguish four different time labels. $t_0$ corresponds to the first moment of light emission. $t_1$ is the time of flight between the source and the HSM of the first wave front, during this time the HSM has moved the distance $vt_1$. $t_2$ is the time for the round trip of the longitudinal and transversal wave fronts. During this time the mirrors have moved the distance $vt_2$. Finally, after the time $t_3$, the two waves superposed at point P that has moved the distance $vt_3$. The ideas of the angular effect found in section \ref{misang} must be applied to solve for the direction of the reflected waves, but as we can see the problem becomes much more complicated. B. M. Bolotovskii and S. N. Stolyarov have studied in detail the effects that occurs at moving mirrors as seen from the perspective of the frame in which the mirrors are in motion \cite{bolotovskii}, unfortunately, they treated the problem in the case in which the source is at rest while the mirror is in motion, but they did not investigate the situation when both objects are moving with the same velocity. What is important to mention here is that someone placed at the mirrors would observe aberration. This is schematically shown by the small segments superposed on the vector pointing from the light source to the HSM. These segments indicate the emission of the waves at different instants as the source moves to the right; so that the overall direction of energy flow is slightly tilted to the left. This tilt is, obviously, function of the speed $v$ and the rotational angle $\theta$. After the several reflections that the waves undergo, the corresponding OPL's change in relation to the initial configuration, i.e., $\theta=0$, and thus a fringe shift may appear. Note, however, that if the plane of the interferometer is orthogonal to the earth's motion [see Fig. \ref{introt}(b)] the interferometer does not undergo length contraction. And, provided that we considered the reflection at the mirrors, the problem may be reduced to the case at rest in $S$. In such situation it is not difficult to show that, in spite of the variation of $\theta$, $\delta$ remains constant and therefore no fringe shift is expected. 

\subsection{Imposibility of the measurement of the one-way speed of light}
\label{spedpriv}
What is \emph{quite unjectionable} from the standpoint of any of the theories treated here and that has been, consciously or unconsciously, in the mind of most physicists is that there must exist, at least, \emph{one inertial frame} where the \emph{one-way} speed of light is really isotropic \cite{kennedy,perez,mansouri,guerra2,abreu,iyer1,einstein2,robertson,spavieri,marinov,petry,selleri,levy,puccini}. And I have emphasized ``one-way" in order to distinguish it from the \emph{two-way} speed of light. The trouble is that the one-way speed of light seems to be experimentally inaccessible \cite{zhang,lammerzahl1,lammerzahl2} because all measurements of velocity imply measurements in which the physical entities, in our case electromagnetic fields, have to go in opposite directions, i.e., they have to close a physical circuit (see also \cite{krisher,will}). From the experimental point of view, the key point in a direct measurement of the speed of a physical entity is that, in fact, one really measures the \emph{average, two-way or harmonic mean} and not the absolute or one-way speed (see also \cite{perez,guerra2,abreu,iyer1,guerra3,spavieri,lammerzahl1} and references therein). 

In table \ref{tspeed} we give a summary of the values for the one-way and the two-way speed of light for the frames $S$ and $S'$ under the different theoretical frameworks treated here. In the results presented we have assumed that the light source is co-moving with the frame $S'$ whose speed is $\mathbf{v}$ relative to $S$. For this reason, as seen from $S$, ET disagree with the other theories. Evidently, if the system $S'$ were at rest relative to $S$ all theories would render the one-way and two-way speed of light equal to $c$.
 \begin{table*}[htp]
   \caption{Predictions for the one-way and two-way speed of light in the frames $S$ and $S'$ under the four theoretical frameworks treated in this work: Old Aether Theory (OET), Emission Theories (ET), Special Relativity (SR) and New Aether Theory (NET).}
  \centering 
  \begin{tabular}{c|ccccc}
  \hline
\textbf{Theory:} &{\bf } &  {\bf OET}&  {\bf  ET} & {\bf SR} &  {\bf NET}
\\  \hline \hline 
 Frame $S$ &&&&
  \\ \hline
One-way && $c$ & $|c\mathbf{\hat{k}}\pm \mathbf{v}|$ & $c$ & $c$ 
\\ \\
Two-way & &$c$ & $\frac{2(|c\mathbf{\hat{k}}+\mathbf{v}|)(|c\mathbf{\hat{k}}-\mathbf{v}|)}{|c\mathbf{\hat{k}}+\mathbf{v}|+|c\mathbf{\hat{k}}-\mathbf{v}|}$ & $c$ & $c$ \\ \hline \hline 
 Frame $S'$ &&&&
   \\\hline
 One-way & & $(c\pm \mathbf{\hat{m}'}\cdot \mathbf{v})$ & $c$ & $c$ & $c^2(c\pm \mathbf{\hat{n}'}\cdot \mathbf{v})^{-1}$
\\  \\
 Two-way & &$c^{-1}[c^2- (\mathbf{\hat{m}'}\cdot \mathbf{v})^2]$ & $c$ & $c$ & $c$ \\ \hline \hline 
\end{tabular}
\label{tspeed}
\end{table*}

Now that we have understood the basic notions of the NET the reader can verify that this theory also explains the aberration of light, the Fizeau experiment \cite{fizeau1,fizeau3,michelson2}, the Ives-Stilwell experiment \cite{christov3,ives1,ives}, etc. Therefore, we believe that we have reached a point where we are allowed to express that the same amount of kinematical experiments that validates SR, at the same time, also corroborates the NET. And if we have convinced the reader that the NET is consistent, it is time for him to ask why we should accept a theoretical system that in kinematical terms does not aport, at first sight, nothing new. Before we address this question, first, we should convince ourselves of the importance of the aether. I think that the answer could shed light on epistemological terms, and so, it could give us new insights to search for new theoretical horizons and experimental techniques as well.

\section{New physics or new philosophy?}
\label{newphys}

\subsection{Reasons to reject the \ae ther}
\label{whatresas}
Still there is one central question that requires a brief discussion: Can we conclude from the allegedly  negative result of the experiment that the aether does not exist? After our previous discussion the natural answer would be in the negative. However, some physicists have concluded that there is no aether. They have argued that in the OET if the eather existed as a privileged frame the observer in $S'$ would have observed a shift of fringes as a consequence of the magnitude of the anisotropy of the one-way speed of light. We think that this argument is weak and it has been readily elucidated above. Even after the discovery of the Lorentz transformations most physicists used to associate the aether with an absolute frame and almost instinctively applied GR to analyze a problem \footnote{A similar example of this point can be found in J. D. Jackson \cite{jackson} pp. 519-522.}. This procedure is misleading; first, we should bear in mind that physicists embraced the Galilean transformations in the OET because they were the only transformations known at that time. And, secondly, that the aether as physical reality is not matter of metaphysics but of logic. As a science, physics must be coherent not only mathematically but also physically. And nature is telling us so emphatically that \emph{all classical waves require a material medium}. But when we refused the aether we not only deprived the universe of a privileged frame but also of mass. Instead, the Maxwellian aether was replaced by the gravitational aether (four-dimensional space-time) \cite{einstein10} which obviously lacks the whole mass of the original one since the stress-energy tensor can only account for the mass of baryonic matter, leptons, gravitational binding energy and radiation. Then, if there is no aether, this rises the question: where do dark matter come from? 

\subsection{Aspects of the \ae ther}
As regards to the nature of the aether, and no less to that of dark matter \cite{beltran,feng}, up to now there is no satisfactory model. The ancient fluid and elastic-solid models \cite{max1,whittaker,larmor1,larmor2} were unfruitful attempts and after the introduction of the SR and the GTR the theoretical research of the aether properties halted. Therefore, the issue remains open and, in this sense, it is plausible to propound a unified theory that links the aether with dark matter. This connection is epistemologically possible because, for us, space is made up of matter whose intrinsic properties are different from those of ponderable matter, i.e., baryonic and leptonic matter. Therefore, space seen as a material medium might unravel cosmological puzzles like the horizon problem, the fly-by and the pioneer anomalies \cite{petry1}, dark energy \cite{christov} and explain, on equal footing to the general theory of relativity, the perihelion of mercury, the bending of light rays, the gravitational lensing, etc. In this respect Ye Hing-Hao and Lin Qiang \cite{hao} have exposed the similarities between the light propagation in a curved spacetime and that in a medium with graded refractive index, they remark that ``a curved spacetime is equivalent to an inhomogeneous vacuum". At the microscale, Volovik has shown that ``The inhomogeneous deformations of the condensed matter ground state --quantum vacuum-- induce nontrivial efective metrics of the space, where the free quasiparticles move along geodesics, thus simulating the gravity field". Consoli has also pointed out that the vacuum condensates, which are generally accepted in elementary particle physics, can be seen, in fact, as empty space (Newtonian space) filled of Higgs particles whose properties may have a direct connection with the aether notion \cite{consoli1,consoli3}. Furthermore, Cahill and Kitto \cite{cahill} have presented experimental evidence that relates the aether with the cosmic background radiation (CBR). Following this idea there exists the work of Mansouri and Sexl \cite{mansouri} who have considered the CBR as a privileged frame to develop the test theories of SR. Examples of this sort suggest one simple word: \emph{Unification}. Dark matter, the vacuum and the CBR could be aspects of the same thing. Work towards the unification of these concepts and, more important, towards the creation of a simplified model of physical reality is under way. String theory and loop quantum gravity are admirable efforts for a unified theory, we are convinced, however, that, similar to the case of the geocentric model, at the end, a profound change in the perception of reality will cost less. This conversion has been recently worked out by C. I. Christov under the theory of wave mechanics, from which it is shown that ``the governing equations of an incompressible elastic continuum yield Maxwell's equations as corollaries'' \cite{christov,christov2,dmitriyev}. Under this theory, electromagnetic waves are, in fact, shear waves of the material continuum (some other similar works can be found elsewhere \cite{thornhill,funaro,su,wang}). He has also been able to construct a model that not only unifies gravitation, electrodynamics and quantum wave mechanics but also explains the wave-particle duality \cite{christov1}. 

Finally, I would like to add one last remark regarding relativistic effects. In the previous sections we held the position that length contraction and time dilation are actual effects. Since we all know that our measuring instruments are made up of matter it is natural to wonder about the physical mechanisms that matter undergoes to show length contraction. A satisfactory description of reality cannot be found in a purely geometrical theory and thus neither SR, nor the NET, nor the GTR would suffice for this purpose. In these respects, Christov has also shown that particles can be seen as phase patterns (derived from the notion of solitons or quasiparticles) that propagate over the material continuum. As the phase patterns move they undergo contraction in the direction of motion in proportion to the Lorentz factor \cite{christov,christov2,christov1}.

\section{Conclusions}
\label{con}
The present article contains original material that we believe is crucial for the understanding of nature. We claim to give a comprehensive review of the subject, our goal being essentially to dispel some misconceptions around the experiment and to establish a clear distinction between the physical concepts (baggage) and its relation with the mathematical laws. We have stressed the most significant features of the theories and it has made clear that there are no strong arguments to refuse the aether. Therefore, if we accept it as part of the universe then we could have a physical background to justify the unresolved puzzles, e.g. dark matter.
 
To end this contribution I would like to quote what Maxwell wrote on the last page of his treatise  \cite{max3}:

\begin{description}
  \item[ ] \textsf{...whenever energy is transmitted from one body to another in time, there must be a medium or substance in which the energy exists after it leaves the body and before it reaches the other, for energy, as Torricelli remarked, `is a quintessence of so subtile a nature that it cannot be contained in any vessel except the inmost substance of material things.' Hence all these theories [Betti, Neumann, Riemann, etc.]\footnote{Words in brackets were introduced by the author of this memoir.} lead to the conception of a medium in which the propagation takes place, and if we admit this medium as an hypothesis, I think it ought to occupy a prominent place in our investigations, and that we ought to endeavour to construct a mental representation of all the details of its action, and this has been my constant aim in this treatise.}
\end{description}

This paragraph, I believe, reflects our present reality. We are still inquiring the properties of such subtle matter which any purported final theory must have to address. In a forthcoming contribution we shall give a detail exposition of the physics behind gravity which shall be treated as a flow of that subtle matter. Such theory will make more intelligible \emph{the physics surrounding the Michelson-Morley experiment}.

\subsection*{Acknowledgments}
The author is grateful to Georgina Carrillo and Dr. Luis Ure\~na for invaluable comments. A CONACYT grant is acknowledged. This memoir is to respectfully honor the memory of the Maxwellians and those who loyally believe in the \AE ether: Ren\'e Descartes, Isaac Newton, Christiaan Huyghens, George Airy, Fran\c{c}ois Arago, Augustin Fresnel, Thomas Young, George Green, Michael Faraday, William Thomson, John Henri Poynting, Hermann von Helmholtz, Heinrich Hertz, Joseph Larmor, Hendrik Antoon Lorentz, Henri Poincar\'e, Albert Michelson, Dayton Miller, Edmund T. Whittaker, Joseph J. Thomson, C. K. Thornhill, et alii.

\end{document}